\documentclass[]{pasj01}
\usepackage{xcolor,ulem,url, lscape}
\usepackage{longtable}
\usepackage{lscape}
\usepackage{dcolumn}
\usepackage{comment}
\usepackage{cuted}
\usepackage[switch,mathlines]{lineno}

\Received{$\langle$reception date$\rangle$}
\Accepted{$\langle$acception date$\rangle$}
\Published{$\langle$publication date$\rangle$}

\begin{document}

\title{
An ultra wide-band, high-sensitivity Q-band receiver for single-dish telescopes, eQ: 
rest frequency determination of CCS ($J_N$ = $4_3$--$3_2$) and SO ($J_N$ = $1_0$--$0_1$), and high-redshift CO ($J$ = 1--0) detection
}
\author{Fumitaka \textsc{Nakamura}\altaffilmark{1,2,3},
Chau-Ching \textsc{Chiong}\altaffilmark{4}, Kotomi \textsc{Taniguchi}\altaffilmark{1},  Chen \textsc{Chien}\altaffilmark{4}, Chin-Ting \textsc{Ho}\altaffilmark{4}, Yuh-Jing \textsc{Hwang}\altaffilmark{4}, You-Ting \textsc{Yeh}\altaffilmark{4,5}, Tomomi  \textsc{Shimoikura}\altaffilmark{6},  Yasumasa \textsc{Yamasaki}\altaffilmark{7}, Sheng-Yuan \textsc{Liu}\altaffilmark{4}, 
Naomi \textsc{Hirano}\altaffilmark{4},
 Shih-Ping \textsc{Lai}\altaffilmark{5},  Atsushi \textsc{Nishimura}\altaffilmark{8}, Ryohei \textsc{Kawabe}\altaffilmark{1}, Kazuhito \textsc{Dobashi}\altaffilmark{9},
  Yasunori \textsc{Fujii}\altaffilmark{1}
 Yoshinori \textsc{Yonekura}\altaffilmark{10},
Hideo \textsc{Ogawa}\altaffilmark{7}, and 
Quang \textsc{Nguyen-Luong}\altaffilmark{11}
}%
\altaffiltext{1}{National Astronomical Observatory of Japan, 2-21-1 Osawa, Mitaka, Tokyo 181-8588, Japan}
\altaffiltext{2}{The Graduate University for Advanced Studies
(SOKENDAI), 2-21-1 Osawa, Mitaka, Tokyo 181-0015, Japan}
\altaffiltext{3}{Department of Astronomy, The University of Tokyo, Hongo, Tokyo 113-0033, Japan}
\altaffiltext{4}{Institute of Astronomy and Astrophysics, Academia Sinica, 11F of Astronomy-Mathematics Building, National Taiwan University, No 1, Sec., Roosevelt Road, Taipei, 10617 Taiwan}
\altaffiltext{5}{Institute of Astronomy, National Tsing Hua University, No. 101, Section 2, Kuang-Fu Road, Hsinchu 30013, Taiwan}
\altaffiltext{6}{Faculty of Social Information Studies, Otsuma Women’s University, Chiyoda-ku,Tokyo, 102-8357, Japan}
\altaffiltext{7}{Department of Physics, Graduate School of Science, Osaka Metropolitan University, 1-1 Gakuen-cho, Naka-ku, Sakai, Osaka 599-8531, Japan}
\altaffiltext{8}{Nobeyama Radio Observatory, National Astronomical Observatory of Japan (NAOJ), National Institutes of Natural Sciences (NINS), 462-2 Nobeyama, Minamimaki, Minamisaku, Nagano 384-1305, Japan}
\altaffiltext{9}{Department of Astronomy and Earth Sciences, Tokyo Gakugei University, 4-1-1 Nukuikitamachi, Koganei, Tokyo 184-8501, Japan}
\altaffiltext{10}{Center for Astronomy, Ibaraki University, 2-1-1 Bunkyo, Mito, Ibaraki 310-8512, Japan}
\altaffiltext{11}{Department of Computer Science, Mathematics \& Environmental Science, American University of Paris PL111, 2 bis, passage Landrieu, 75007 Paris, France}
\email{fumitaka.nakamura@nao.ac.jp}


\maketitle

\begin{abstract}
We report on the development and commissioning of a new Q-band receiver for the Nobeyama 45-m telescope, 
covering 30--50 GHz with a receiver noise temperature of about 15 K. We name it eQ (extended Q-band) receiver.
The system noise temperatures for observations are measured to be $\sim$ 30 K at 33 GHz and $\sim$ 75 K at 45 GHz.
The Half-Power-Beam-Width (HPBW) is around 38\arcsec at 43 GHz.
To enhance the observation capability, we tested the smoothed bandpass calibration technique and demonstrated the observation time can be significantly reduced 
compared to the standard position switch technique.
The wide-bandwidth capability of this receiver provides precise determination of rest frequencies for molecular transitions with an accuracy of a few kHz through simultaneous observations of multiple transitions. Particularly, we determined the rest frequency of SO ($J_N$ = $1_0$--$0_1$) to be 30.001542 GHz, along with the rest frequency of CCS ($J_N$ = $4_3$--$3_2$) being 45.379033 GHz, adopting CCS ($J_N$ = $3_2$--$2_1$) at 33.751370 GHz as a reference line.
The SO profile shows a double peak shape at the Cyanopolyyne Peak (CP) position of the Taurus Molecular Cloud-1 (TMC-1).
The SO peaks coincide well with the CCS sub-components located near the outer parts of the TMC-1 filament. We interpret that the gravitational infall of TMC-1 generates shocks which enhance the SO abundance.
The TMC-1 map shows that carbon-chain molecules are more abundant in the southern part of the filament, whereas SO is more abundant in the northern part.  
The eQ's excellent sensitivity allowed us to detect faint CO ($J$ = 1--0) spectra from the high-redshift object at a redshift of 2.442.
Our receiver is expected to open new avenues for high-sensitivity molecular line observations in the Q-band.
\end{abstract}

\section{Introduction}
\label{sec:intro}

Low-frequency observations such as Q-band (conventionally, 33--50 GHz) require a large single-dish telescope or interferometers to spatially resolve molecular cloud structures. 
The Nobeyama Radio Observatory (NRO) 45-m telescope is one of the largest single-dish telescopes and is a crucial instrument for such observations.
Currently, the Nobeyama 45m telescope has two Q-band receivers, H40 and Z45.
The H40 is the telescope reference receiver, often used for pointing observations with SiO masers from evolved stars at about 43 GHz. It is a single-linear polarization receiver covering the RF frequency of 42.5--44.5 GHz.
The Z45 is a dual-linear polarization receiver covering 42--46 GHz \citep{nakamura15}. 
It was designed to conduct the Zeeman observations with 45 GHz CCS $J_N$ = $4_3$--$3_2$ \citep{nakamura19}. It has also been used for other standard observations such as position-switch one-point observations \citep{taniguchi16,taniguchi18,dobashi18} and on-the-fly (OTF) mapping observations \citep{nakamura14,dobashi19}. 
However, the narrow bandwidth of the existing Q-band receivers on the 45-m telescope limits their usefulness for standard molecular line observations. Particularly, the 50 K receiver noise temperature of Z45 is too high for conducting multi-position Zeeman observations in a limited observation time.

To overcome this limitation, we have developed a new Q-band dual-linear polarization receiver called eQ (extended Q-band), and installed it on the Nobeyama 45-m telescope \citep{chiong22}. 
{\color{black} The eQ receiver covers an RF bandwidth from 30 GHz to 50 GHz (hereafter, we refer this range to the Q-band or extended Q-band), making it the widest bandwidth Q-band receiver installed on any large single-dish telescope for astronomical science observations.
Furthermore, with its receiver noise temperature about 15 K, which is much lower than that of the Z45 receiver (50 K), the eQ receiver also offers a better sensitivity than the existing receivers over the RF frequency range.
}

Through the observations of the lines in the Q-band, we primarily focuses on the following scientific objectives for the eQ receiver:
\begin{itemize}
\item[1.] {\bf Zeeman Observations:} 
The eQ receiver, with its ultra-wide frequency coverage, has the unique capability for simultaneous Zeeman observations of the multiple molecular transitions having large 
Land\'{e} factors, which are the most powerful and direct method for deriving magnetic field strengths in dense interstellar cores.
We aim to conduct Zeeman observations in the following transitions, including CCS (thioxoethenylidene) in its $J_N$ = $4_3$--$3_2$ ($\sim$ 45 GHz) and $J_N$ = $3_2$--$2_1$ ($\sim$ 34 GHz) transitions, and SO (sulfur monoxide) in its $J_N$ = $1_0$--$0_1$ ($\sim$ 30 GHz) transition.
These molecules have unpaired electrons, which results in relatively large spectral line splitting in the presence of a magnetic field. 
\item[2.] {\bf High-Redshift Molecular Line Detection:} 
{\color{black} 
The Q-band is best matched with the lowest-rotational transitions, $J$ =1--0, and $J$ = 2--1, of CO (carbon monoxide) at redshift $z$ $\sim$ 1--7, the cosmic epoch of most active galaxy formation.
By detecting these CO lines from high-redshift objects, we aim to estimate the total molecular gas content and properties.}

\item[3.] {\bf Astrochemistry Exploration in the 30--50 GHz Band:} We also aim to explore astrochemistry in the Q-band (30--50 GHz) frequency range. In this band, there are various interesting molecular lines that can provide valuable insights into the chemistry of the interstellar medium. This includes the study of different molecular species and their abundances, helping to advance our understanding of chemical processes in space.
\end{itemize}

In this paper, we present the receiver specifications and the results of science commissioning observations. 
The details of the receiver specifications can be found in Section \ref{sec:spec}, 
while \citet{chiong22} provides a more comprehensive overview of these specifications.
In section \ref{sec:line}, we briefly describe selected molecular lines in the Q-band.
In Section \ref{sec:sbc}, we highlight the effectiveness of the Smoothed Bandpass Calibration (SBC) technique for eQ position-switch observations, demonstrating its ability to significantly reduce the total observation time.
Furthermore, Sections \ref{sec:obs1} and \ref{sec:obs2} focus on the initial results obtained during the commissioning observations. Particularly, we have conducted measurements on several molecular lines with rest frequencies that do not match well across various catalogs. Benefiting from the wide bandwidth of our receiver, we were able to simultaneously observe these lines, which allows us to directly compare the lines and minimize any differences or errors associated with the rest frequencies in the catalogs. 
Using these results, we briefly discuss the kinematics of the nearby dense filament, Taurus Molecular Cloud-1 (TMC-1).
Lastly, we summarize our results in Section \ref{sec:summary}.

\begin{center}
 \begin{table*}
  \caption{eQ receiver specifications}
 \begin{tabular}{lcc}
 \hline
 Parameter & Value & Note \\ \hline
 Type of receiver & Cooled HEMT (2SB) &  \\ 
 Optics system & cold &  \\ 
 RF frequency coverage (GHz)& 30--50 &   \\ 
 Local Oscillator & Signal generator with frequency tripler & \\ 
 IF frequency range (GHz) & 4--8 & \\
LO frequency range (GHz) & 34--42 & \\
Polarization & Dual linear polarization &  \\ 
Receiver noise temperature & $\sim$ 15 K &  \\ 
System noise temperature & 70--90 K at 45 GHz& Excellent sky condition \\ 
 & 30--60 K at 33 GHz& Excellent sky condition  \\ 
FWHM main beam size  & 38.8 \arcsec $\times$36.6\arcsec  at 43 GHz& SiO masers (NML Tau), 2023--2024 season \\ 
Aperture beam efficiency ($\eta_A$) & 0.61/0.61 at 31/43 GHz  & Jupiter\\ 
Main beam efficiency ($\eta_m$)  & 0.75/0.73 at 31/43 GHz  & Jupiter\\ 
   \hline
 \end{tabular}
 \label{tab:eQ}
\end{table*}
\end{center}

\begin{figure*}[tbp]
 \begin{center}
  \includegraphics[width=\linewidth]{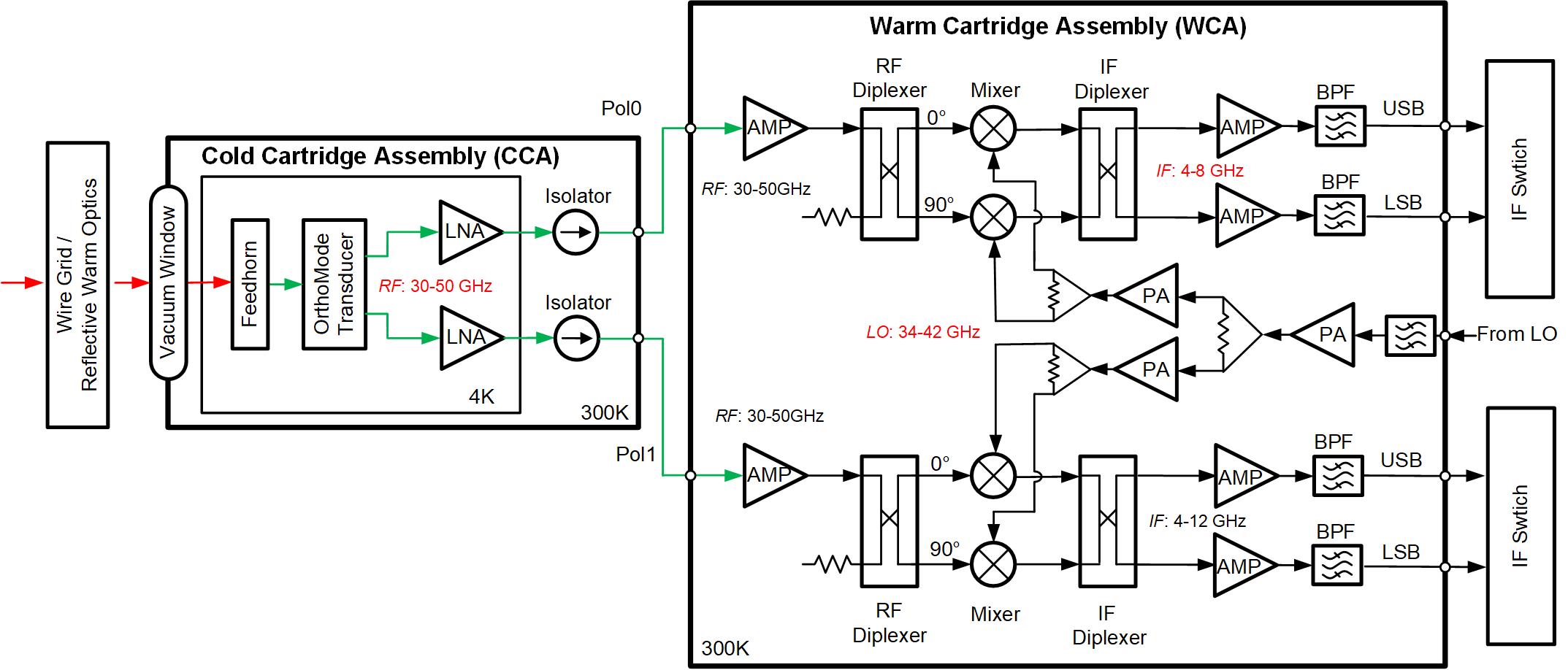}
 \end{center}
\caption{Block diagram of the eQ receiver. The arrows denote the signal path from the telescope to the IF processor. Green arrows stand for waveguide connection, while black ones are for coaxial connection. Red arrows represent propagation in free space.}
\label{fig:blockdiagram}
\end{figure*}

\section{The eQ receiver parameters}
\label{sec:spec}

\subsection{eQ architecture}

\begin{figure*}[htbp]
 \begin{center}
  \includegraphics[width=7.5cm]{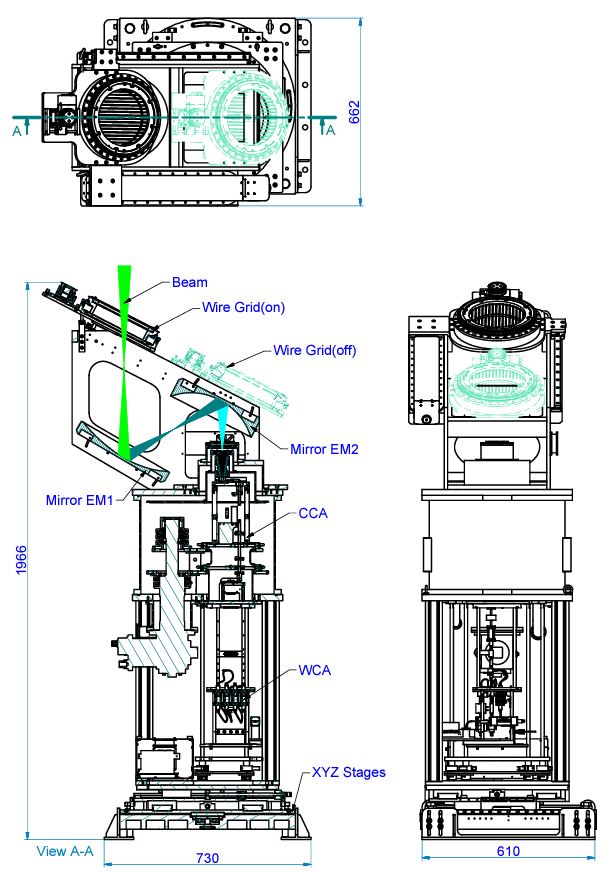} \hspace*{0.5cm} 
 \includegraphics[width=7.5cm]{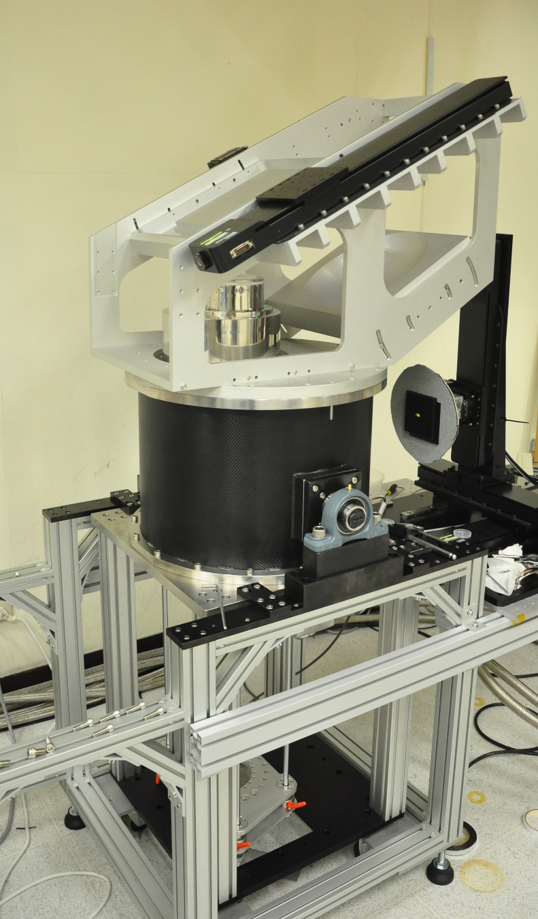}
 \end{center}
\caption{The eQ receiver layout ({\it left}) and  the image of the receiver during the lab testing ({\it right}). The incoming signal trajectory is also shown. Two ellipsoidal mirrors (EM1, EM2) are used to focus the beam. The wire grid on top is for the purpose of performing polarization calibration. After the calibration is finished, it can be moved to "off" position remotely for normal observation.}
\label{fig:layout}
\end{figure*}

\begin{figure}
 \begin{center}
  \includegraphics[width=7cm]{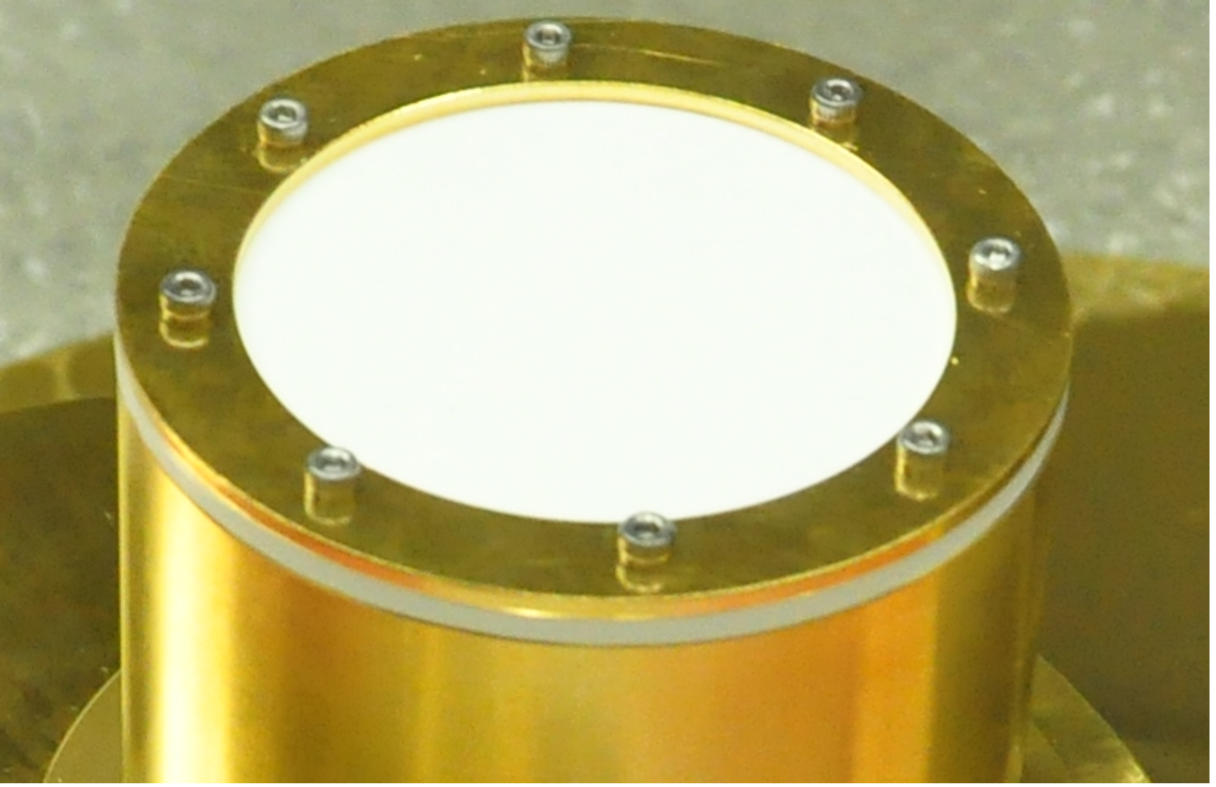} 
 \end{center}
\caption{Zitex IR filter between the vacuum PTFE window and feerhorn}
\label{fig:if_filter}
\end{figure}

The receiver architecture, as illustrated in Figure \ref{fig:blockdiagram}, follows the fundamental design of the ASIAA-developed ALMA Band-1 receiver \citep{hwang17}. 
Receiver production was led primarily by the ASIAA team.
The size of the eQ receiver is 730 mm $\times$ 620 mm $\times$ 1900 mm and weight is about 440 kg
(see Figure \ref{fig:layout}).  
It comprises two main components: the cold cartridge assembly (CCA) and the warm cartridge assembly (WCA). 
For further details about the receiver, refer to \citet{chiong22}. 

CCA consists of two stages operating at 4K and 50K. Both stages are supported by G10 glass fiber tube spacers, which have very low thermal conductivity. All the components are assembled to the 4K stage, with two long WR22 output waveguide attached to the bottom room temperature plate. The 50 K stage is used to provide thermal shielding to the 4K stage. When the stainless waveguide passes through the 50K stage shielding, oxygen-free copper strips are attached to relieve conductive heat load from the room temperature stage. On the other hand, the vacuum window towards the telescope is made of polytetrafluoroethylene (PTFE). The radiative heat transmission from the window is blocked by Zitex IR filter between the window and the feedhorn (see Figure \ref{fig:if_filter}).

\begin{figure*}
 \begin{center}
  \includegraphics[width=5cm]{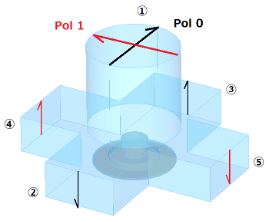} 
    \includegraphics[width=5cm]{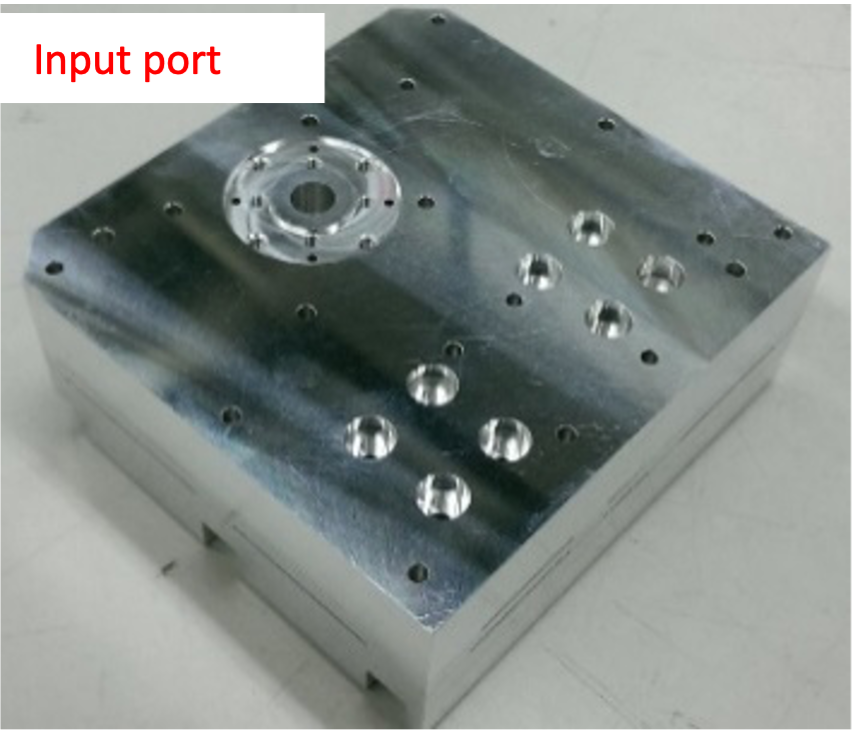} 
      \includegraphics[width=5cm]{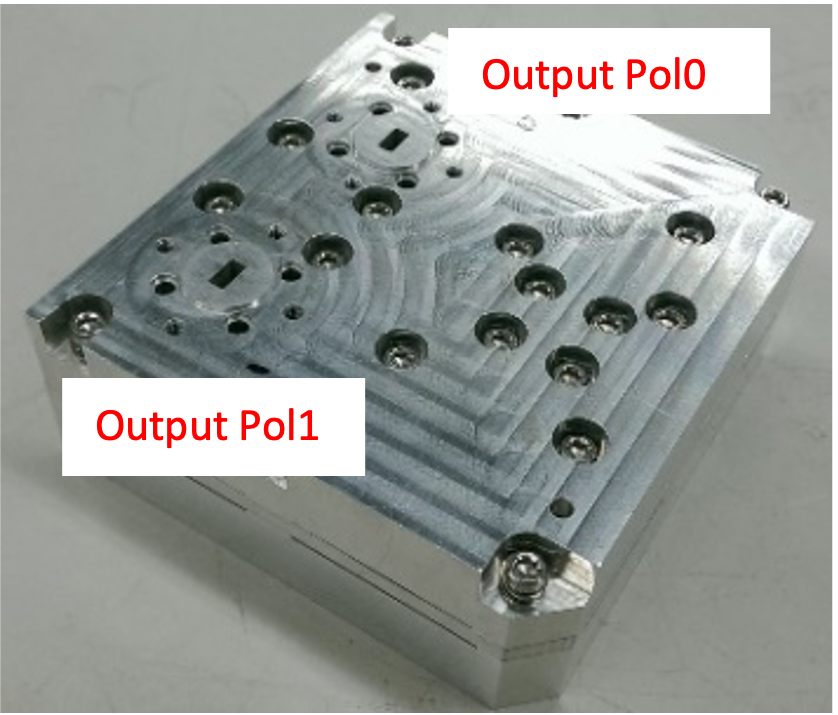} 
 \end{center}
\caption{The newly developed OMT.  The Pol0 ({\it black}) and Pol1 ({\it red}) components are separated as shown in the left panel. The middle and right images indicate the assembled OMT viewing from top (middle) and from bottom (right).}
\label{fig:omt}
\end{figure*}

An outstanding improvement is the receiver's wide bandwidth capability, covering the frequency range of 30 to 50 GHz. This surpasses the bandwidth of the ALMA Band-1 receiver (35--50 GHz). To enhance its sensitivity and minimize the thermal noise, we implemented a cooled optics system and integrated the Low Noise Factory 28--52 GHz cryogenic low noise amplifiers (LNA). The LNA has an excellent noise temperature of 8 K (30--40 GHz) up to 15 K (50 GHz).
At room temperature, the optics system covers 30--50 GHz with a return loss better than 30 dB and a cross-polarization of  $< -30$ dB at band-center and $< -27$ dB at band-edge. For further details of the feedhorn and OMT, refer to \citet{chiong21} and \citet{chiong18}.
The new turnstile junction OMT, which separates the incoming signal into two linearly-polarized components, covers 29--50 GHz with return loss better than 18 dB and isolation better than 40 dB (see Figure \ref{fig:omt} for the images of the OMT). We also developed two ellipsoidal mirrors to guide the radio signals into the feed horn. This configuration minimizes cross-polarization in the optics.

WCA is operating at a room temperature with a two side-band (2SB) down-converter. Here RF signal of 30--50 GHz is down-converted into IF signal of 4--8 GHz with double-balanced mixer MM1-2567LS from Marki Microwave. 2SB is also known as sideband separating scheme for signals from both upper- and lower-sidebands can be obtained simultaneously at two individual output ports as shown in Figure 1.

\begin{figure*}
 \begin{center}
  \includegraphics[width=0.45 \linewidth]{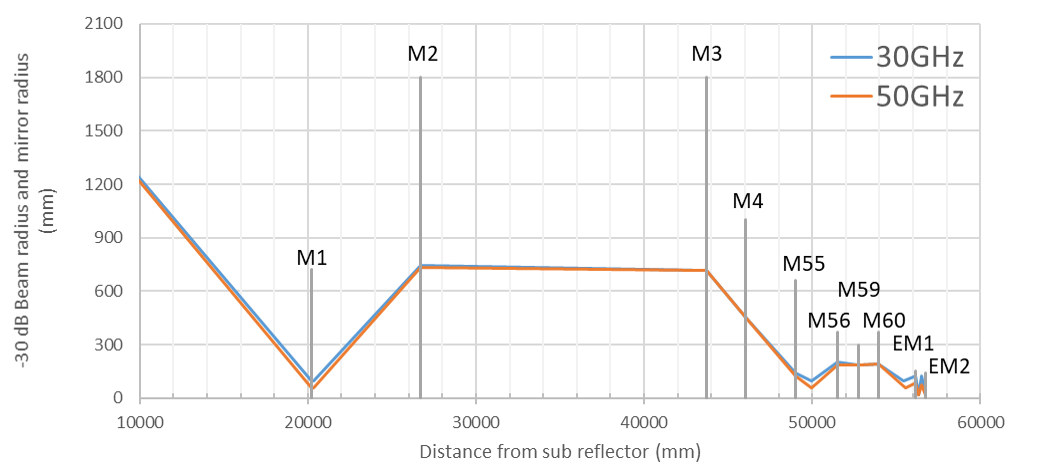}
    \includegraphics[width=0.45 \linewidth]{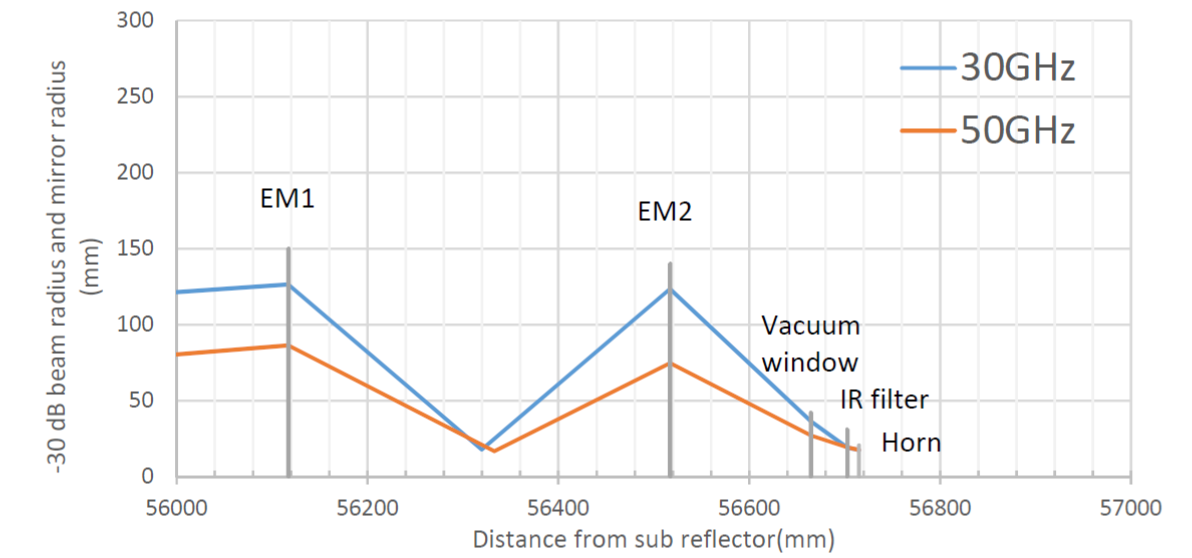}
 \end{center}
\caption{The eQ beam edge level at NRO system. M1 through M60 are the names of the mirrors used in the Nobeyama 45-m beam transmission system. EM1 and EM2 are the ellipsoidal mirrors developed for the eQ.}
\label{fig:mirror}
\end{figure*}

In Figure \ref{fig:mirror} we present the beam edge levels in the 45m telescope system. The beam radius at $-$30 dB power level relative to the beam center and the mirror radius of the eQ receiver are plotted at 30 GHz and 50 GHz, respectively. These figures show that the eQ beam is fitted well in the Nobeyama 45-m beam transmission system.

\begin{figure}[htbp]
 \begin{center}
  \includegraphics[width=\linewidth]{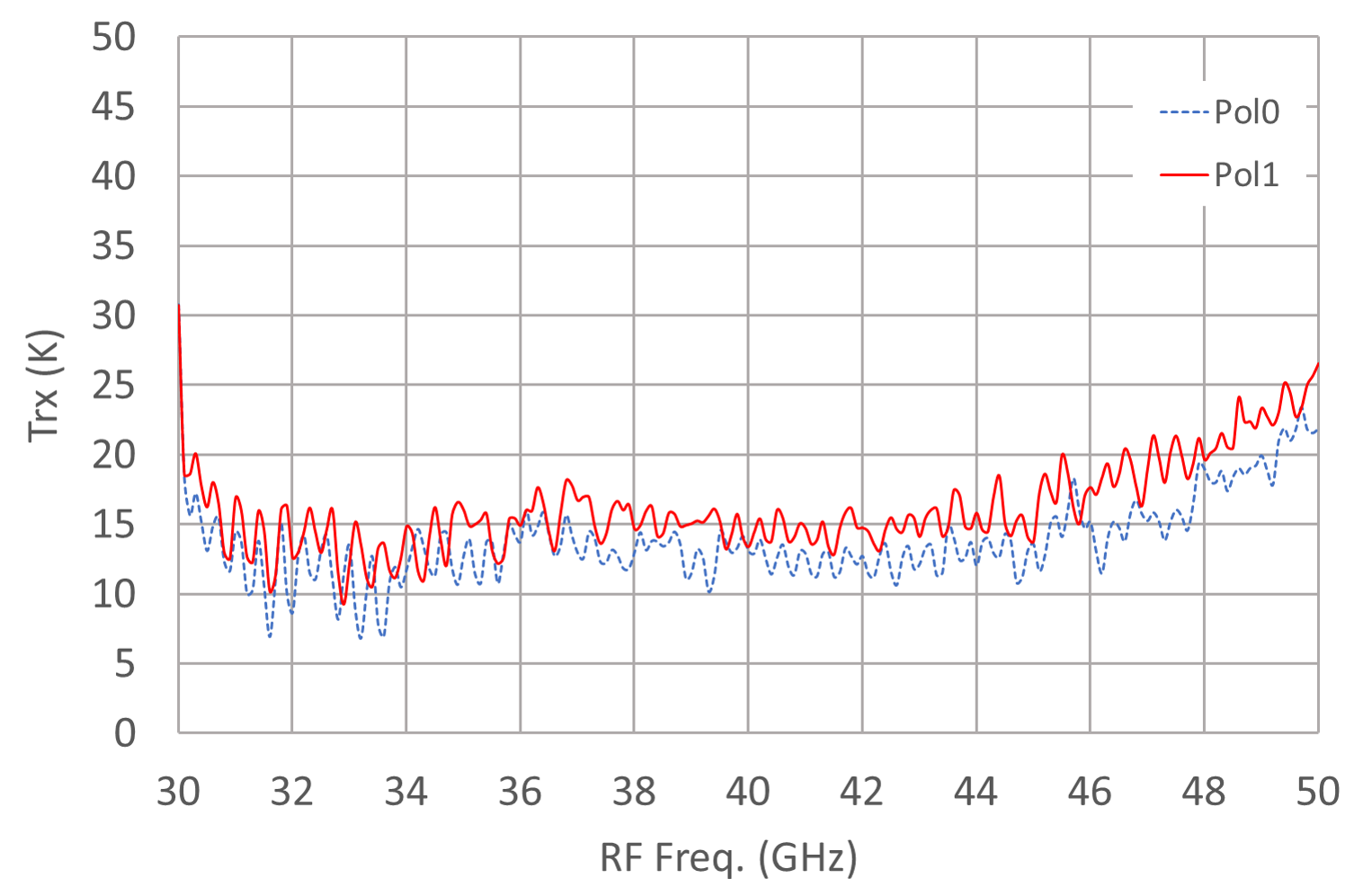}
 \end{center}
\caption{Receiver noise temperature measured in the laboratory. The red and blue lines indicate the receiver noise temperatures for Pol 1 and Pol 0, respectively.}
\label{fig:Trx}
\end{figure}

\subsection{Laboratory measurements}

Laboratory tests were conducted at ASIAA to evaluate the receiver's performance. 
The measurement was performed by a YIG bandpass filter with a power meter. The power level at each stage in the receiver was carefully controlled so that it contributes less than 1\% power compression. The whole system power compression level is less than 2\%.
The measured receiver noise temperature ($T_{\rm rx}$) is presented in Figure \ref{fig:Trx}. The receiver noise temperature was derived by measuring hot (room temperature, $T_{\rm hot}$) and cold (liquid nitrogen, $T_{\rm cold}$) loads 
using the Y-factor technique as, 
\begin{equation}
 T_{\rm rx} = {T_{\rm hot} - 
 Y T_{\rm cold} \over Y-1}\ .
\end{equation}
Throughout the measurement process, the cold load temperature was regularly calibrated and assumed to be 81.5 K for the calculations in Figure \ref{fig:Trx}.
The error in the Y-factor calculation is dominated by the following two items:
1) Reading accuracy of hot and cold load equivalent temperature: $\pm$ 0.3 \% 
in $T_{\rm hot}$, $T_{\rm cold}$ 
with DT-470 temperature sensor according to the Lakeshore manual, 
2) Reading accuracy of the power meter and the power meter linearity: 1\% estimated for the Keysight EPM series power meter with N8480 series power sensor.

Error on $T_{\rm rx}$ can be estimated as
\begin{eqnarray}
    {\Delta T_{\rm rx} \over T_{\rm rx}}
   & = & \left[
    \left({\Delta T_{\rm hot} \over Y-1} \right)^2
    + \left({Y \Delta T_{\rm cold} \over Y-1} \right)^2  \right. \nonumber \\
    &+& \left. \left({(-T_{\rm cold} (Y-1)- T_{\rm hot}+T_{\rm cold}  Y )\Delta Y \over 
    \left({Y-1} \right)^2}\right)^2
    \right]^{1/2}
\end{eqnarray}
For this receiver, $T_{\rm rx} \sim $ 15 K, and therefore we expect $\sim$ 2 K error.

The cold load temperature is calibrated by placing an AN72 Eccosorb absorber cone covering onto the viewing window, with and without liquid nitrogen. By assuming that the receiver sees an ideal 77.35 K load with liquid nitrogen and power reading, $P_{\rm cold-cone}$, together with the ambient load at room temperature with power reading, $P_{\rm hot-cone}$, the equivalent load temperature of the liquid nitrogen tank ($T_{\rm ref}$) in the measurement is derived by the power reading when the receiver is seeing the liquid nitrogen tank ($P_{\rm ref}$). 
\begin{eqnarray}
    T_{\rm ref} & = & {P_{\rm ref}- P_{\rm cold-cone} \over P_{\rm hot-cone}- P_{\rm cold-cone}}
    (T_{\rm hot-cone}- T_{\rm cold-cone})  \nonumber \\[3mm]
    & + & T_{\rm cold-cone} 
\end{eqnarray}
This is essentially the same procedure as in
the ALMA Band 1 Cartridge Test Procedure.

We note that the increase of noise at the higher frequency side (46--50 GHz) is mainly due to the higher noise of the cryogenic LNA, while the sudden increase at around 30 GHz is due to the feedhorn and OMT.

\begin{figure}[htbp]
 \begin{center}
  \includegraphics[width=\linewidth]{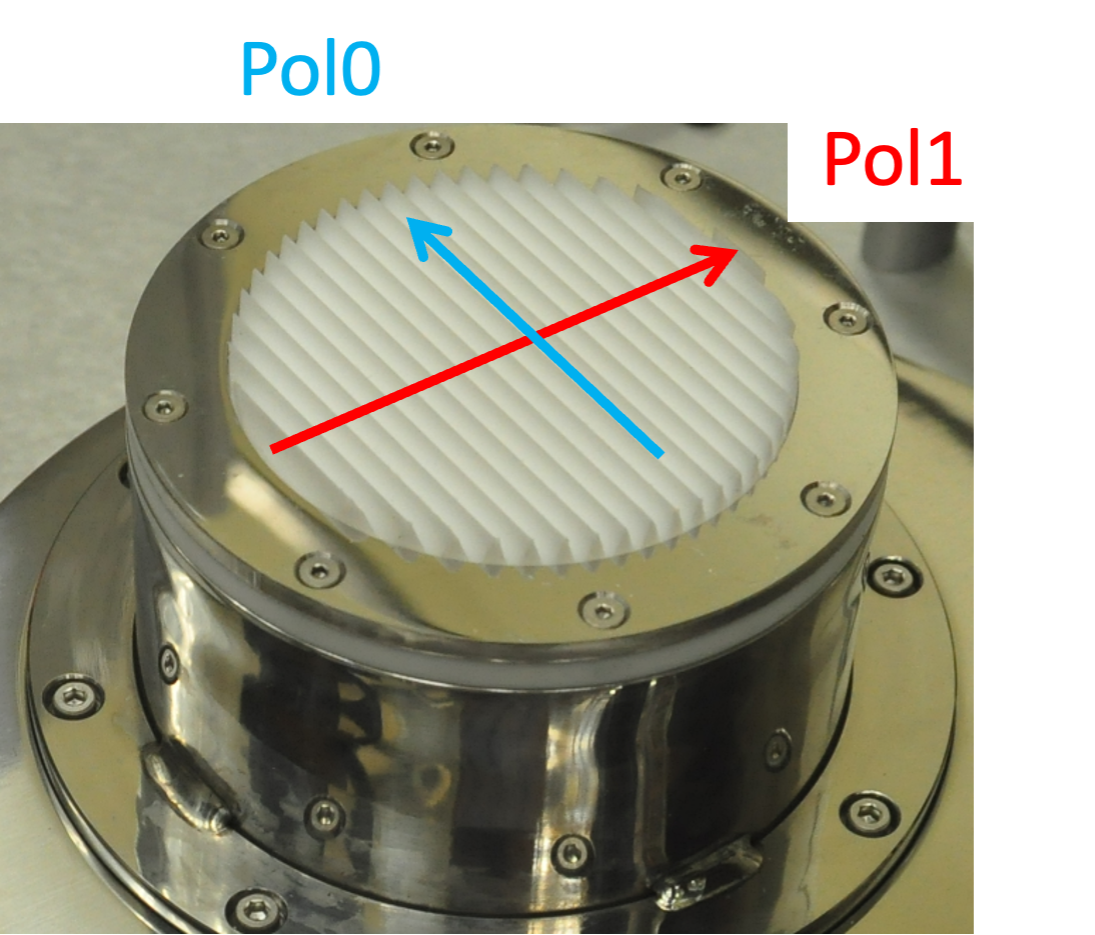}
 \end{center}
\caption{The PTFE grating structure and the directions of two polarized components Pol0 and Pol1.}
\label{fig:grating}
\end{figure}

Two polarization components, Pol0 and Pol1, are defined by the direction of the anti-reflection triangular-shaped grating of the PTFE vacuum window when viewing from the top of the Dewar. Both sides of the vacuum window are with the triangular-shaped grating structure. The side facing the telescope is with the grating parallel to Pol0, and the other side of the vacuum window facing the feedhorn is with the grating parallel to the other polarization (Pol1) for minimizing the difference in reflection coefficients.
Below, we define Pol0 and Pol1 as horizontal and vertical directions, respectively (see Figure \ref{fig:grating}). 

\begin{figure}[htbp]
 \begin{center}
  \includegraphics[width=\linewidth]{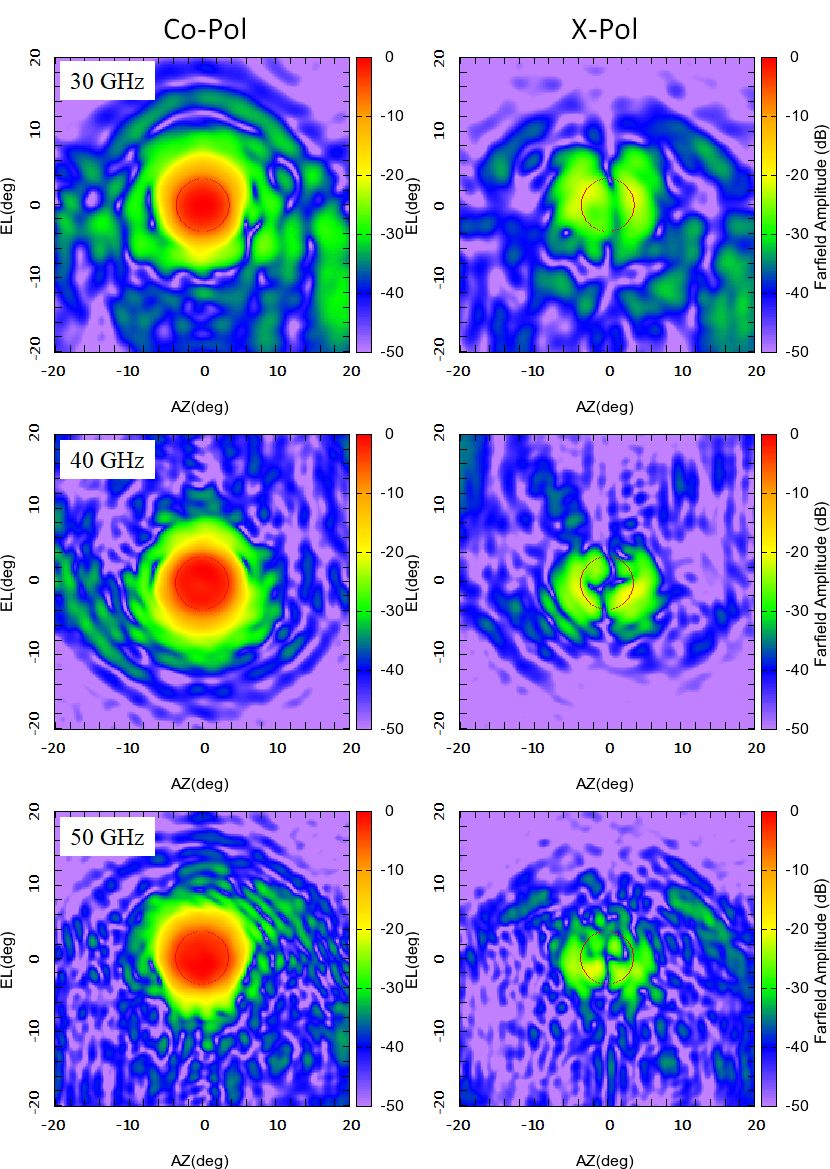}
 \end{center}
\caption{Beam patterns measured in the laboratory.
}
\label{fig:simulated_beam}
\end{figure}

Figure \ref{fig:simulated_beam} displays the beam patterns of the eQ receiver measured in the laboratory at three representative frequencies: 30, 40, and 50 GHz. The measurements were carried out using the Sigma-Koki scanner via the near-field scan technique. The cross-polarization level remained below $-18.5$ dB, while the sidelobe levels were consistently lower than $-23$ dB across all frequencies.

 \begin{table*}
  \caption{Comparison of 30--50 GHz receivers for large single-dish telescopes.}
 \begin{tabular}{lcccccc}
 \hline
  & Green Bank$^1$ &  & Effelsberg$^2$ & Yebes$^3$ & Tianma$^4$ & NRO$^5$ \\ \hline
 Dish diameter & 100m & & 100m & 40m & 65m & 45m \\ 
  $\sigma_{\rm RMS}$ & 240$\mu$m & & 550$\mu$m & 175$\mu$m & 2470$\mu$m & 100$\mu$m \\ 
 $\eta_A$ & 0.58--0.64 & 0.63--0.67 & 0.18--0.27 & 0.59--0.67 & 0.45--0.55 (43 GHz) & 0.50--0.62\\ 
 Receiver name & Q-band & Ka-band & S7mm & Q-band & Q-band  & eQ\\ 
  Frequency (GHz) & 38--50 & 26--40 & 33.5--50 & 31--50 & 35--50  & 30--50\\ 
 Beam  & Two-beam & Two-beam & Two-beam & Single pixel & Two-beam & Single pixel \\ 
Feed & Dual circular & Single linear  & Dual linear & Dual circular & Dual linear & Dual linear \\
$T_{\rm RX}$ & 20--45 K & 10--40 K & --  & 15--40 K & 30--40 K & 13--25 K\\
$T_{\rm sys}$ (Zen.) & 67--160 K & 50--120 K & 110--187 K & 50 K (32.5 GHz) & 80 K (43 GHz) & 30 K (33 GHz)
 \\ 
 &  &  &  & 110 K (48.5 GHz) &  & 75 K (43 GHz)
 \\ 
CCS (45 GHz)$^*$ & $\bigcirc$ & $\times$ & $\bigcirc$  & $\bigcirc$  & $\bigcirc$  & $\bigcirc$ \\ 
CCS (33 GHz)$^*$ & $\times$ & $\bigcirc$ &  $\bigcirc$ & $\bigcirc$ & $\bigcirc$ & $\bigcirc$  \\ 
SO (30 GHz)$^*$  &$\times$ & $\bigcirc$ & $\times$  &$\times$  & $\times$  & $\bigcirc$ \\
   \hline
 \end{tabular}
 \label{tab:comparison}
 $^1$\url{https://www.gb.nrao.edu/GBT/Performance/PlaningObservations.htm} and \url{https://library.nrao.edu/public/memos/gbt/GBT_255.pdf};
 $^2$ \url{https://eff100mwiki.mpifr-bonn.mpg.de/doku.php?id=information_for_astronomers:rx:s7mm_db}; $^3$\citet{tercero21}; $^4$\citet{zhong18}; 
 $^5$this paper and \citet{chiong22};
  $\sigma_{\rm RMS}$: surface accuracy; 
$\eta_A$: aperture efficiency;
$^*$ CCS (45 GHz, $J_N=4_3-3_2$), CCS (33 GHz, $J_N=3_2-2_1$), and SO (30 GHz, , $J_N=1_0-0_1$) are the transitions having large Land\'{e} factors.
\end{table*}

The ripple levels from the 4 IF outputs are better than 4.8 dB (peak-to-peak) for any 2 GHz bandwidth, 1.35 dB (p-p) for any 31 MHz window.
See \citet{chiong21} for more details.

\subsection{Receiver specification}

The receiver was installed in the NRO 45-m cabin in November, 2021.
Measurements of the receiver parameters were conducted during the 2021--2022, 2022--2023, and 2023--2024 observation seasons, yielding consistent results. 
Table \ref{tab:eQ} provides a summary of the receiver specifications.

A brief comparison with the eQ and other Q-band receivers installed in the large single-dish telescopes is presented in Table \ref{tab:comparison}, which is essentially the same as Table 1 of \citet{chiong22}. Table \ref{tab:comparison} indicates
that the eQ exhibits superior sensitivity and bandwidth compared to the listed receivers. 
The receiver noise temperature of the eQ ranges from 13--25 K. 
This is the lowest among the receivers listed in Table 1 of \citet{chiong22}.
In the 2022--2023 season, the best system noise temperatures ($T_{\rm sys}$) were measured to be 30 K for 33 GHz and 75 K for 43 GHz at an elevation of about 70$^\circ$.
Currently, there exist three Q-band receivers that have been recently developed and installed in large single-dish telescopes in the world.
Comparing with the Q-band receiver performance at the Yebes nanocosmo \citep{tercero21} and Tianma (Tian Ma Radio Telescope - TMRT) \citep{zhong18},
the eQ's system noise temperatures are somewhat lower.  In terms of the bandwidth coverage, the Yebes nanocosmo receiver spans from 31.5 to 50 GHz, while the TMRT Q-band receiver covers the range of 35 to 50 GHz. Hence, the eQ receiver stands out with the widest bandwidth and the highest sensitivity among the mentioned Q-band receivers. 

The most notable feature of the eQ is its wide bandwidth, enabling simultaneous observations of the three Zeeman lines: SO ($J_N=1_0-0_1$, 30.00 GHz), CCS ($J_N=3_2-2_1$, 33.75 GHz), and CCS ($J_N=4_3-3_2$, 45.38 GHz). 
The NRO backend, SAM45, offers a frequency resolution of 3.81 kHz with 4096 channels covering 15.625 MHz. This resolution corresponds to velocity resolutions of 0.03 km s$^{-1}$ and 0.025 km s$^{-1}$ at 30 GHz and 45 GHz, respectively, allowing for reasonable distinction of velocity structures. For example, in the TMC-1 filaments, where multiple components with separations of approximately 0.1--0.2 km s$^{-1}$ are overlapped along the line of sight, the finest frequency resolution of SAM45 is essential for resolving such components \citep{dobashi19}.

\begin{figure*}[htbp]
 \begin{center}
  \includegraphics[angle=0, totalheight=7cm]{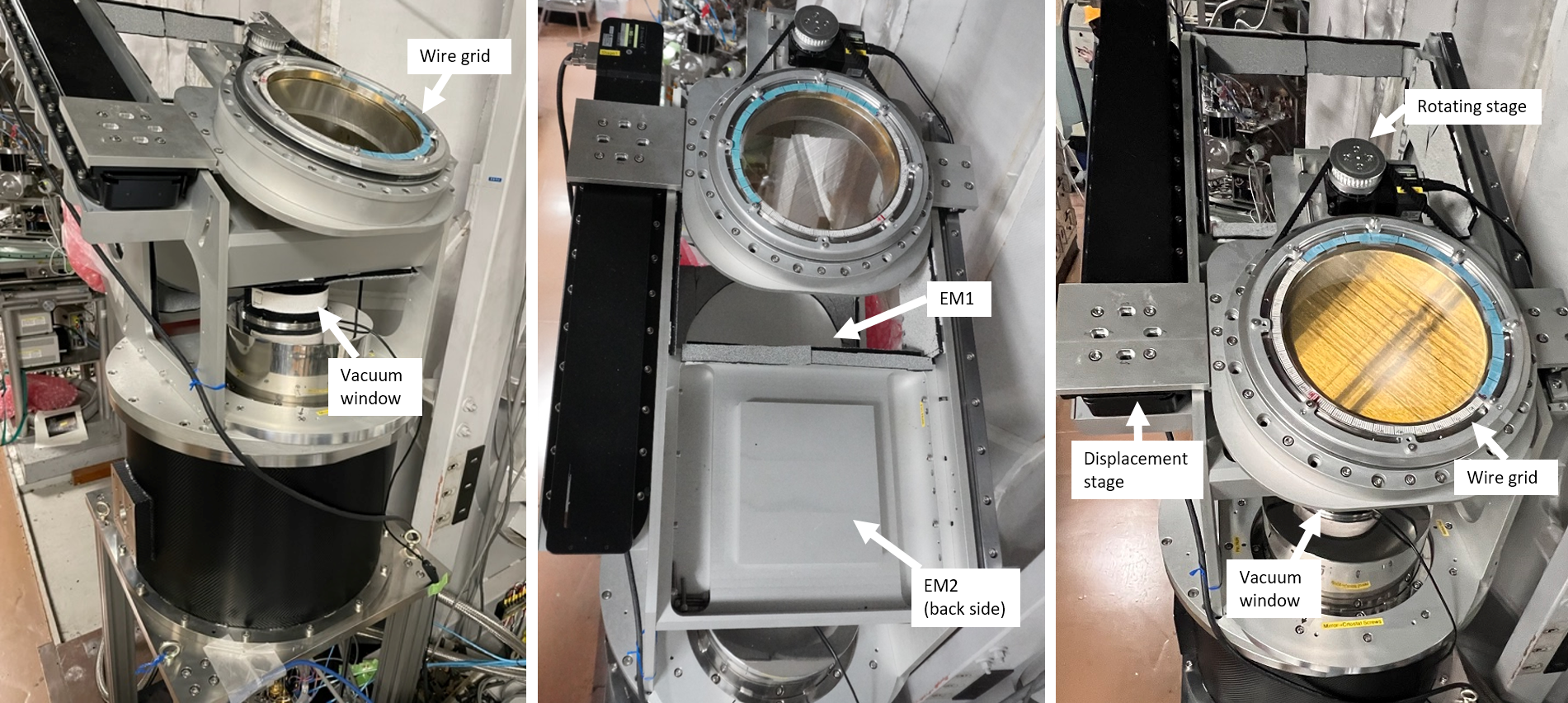}
 \end{center}
\caption{Wire-grid system. ({\it left}) the image of the top of the receiver. ({\it middle}) the wire grid component seen from the top, which is used for acquiring polarization calibration data. This is the on-position. 
({\it right}) same image as the middle panel but for off-position. When we obtain the calibration spectra of the polarized signals with the known directions, the wire grid system is moved remotely to the top of the feedhorn.}
\label{fig:wiregrid}
\end{figure*}

Another distinctive feature of this receiver is its integration of a polarization calibration component located on the top. As depicted in Figure \ref{fig:wiregrid}, the calibration component consists of a circular structure with evenly spaced wires attached. These wires serve to generate linear polarization waves, facilitating the calibration process.
The integration of this specialized component enables the receiver to conduct Zeeman measurements using molecular lines such as CCS and SO. 
With accurate calibrations of the polarization response, the receiver achieves enhanced precision and reliability in polarization measurements.
It is worth noting that the polarization calibration system employed in this receiver is similar to that used in the Z45 receiver.
See \citet{nakamura15}, \citet{nakamura19}, and \citet{mizuno14} for more details of the calibration method. 
This system reinforces the confidence in the receiver's polarization calibration capabilities.

\subsection{Beam pattern}

\begin{figure}[bthp]
 \begin{center}
        \includegraphics[angle=0, width= \linewidth]{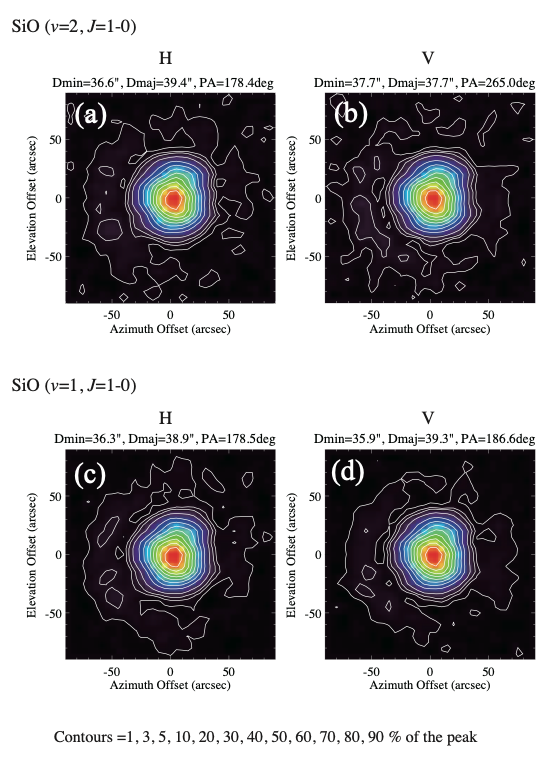}\hspace{1cm}
 \end{center}
\caption{Beam maps of the telescope constructed using the SiO ($v=1$ and $v=2$) maser observations of NML Tau. 
The observations were carried out in the 2023--2024 season.
Panels (a) and (b) illustrate the beam patterns obtained from the horizontal and vertical components of SiO ($v$ = 2, $J$ = 1--0, 42.82057 GHz), respectively. On the other hand, panels (c) and (d) depict the beam patterns derived from the horizontal and vertical components of SiO ($v$ = 1, $J$ = 1--0, 43.12209 GHz), respectively.
The alphabet on the top of each panel indicates the polarization component ({\bf H}orizontal or {\bf V}ertical) being presented. The intensity values are normalized based on the peak value observed.
Contour lines are displayed at the levels of 1, 2, 3, 5, 7, 10, 20, 30, 50, 70, and 90\% of the peak intensity.
Furthermore, the HPBW values, obtained through a two-dimensional Gaussian fit, are provided at the top of each panel.
}
\label{fig:beam}
\end{figure}

Figure \ref{fig:beam} shows the beam patterns of horizontal and vertical polarization components obtained from the  
SiO ($J=1-0$, $v=1$, rest frequency = 43.122027 GHz, and $v=2$, 42.820584 GHz) maser emission of NML Tau (KL Tau). The target coordinates are R.A.= 3h53m28.86s (J2000) and Dec.= 11:24:22.4 (J2000) with $V_{\rm LSR}= 34$ km s$^{-1}$. 
The observation mode employed was on-the-fly (OTF) covering a spatial area of $3\arcmin \times 3\arcmin$ area and a frequency resolution was  set to 30.52 kHz.
Each scan row had
a spacing of 5\arcsec, and the scan time per row was 10 sec. 
We carried out OTF observations in September 1st, 2023. 
We use the NRO digital spectrometer SAM45 as the backend. 
The resulting images of the SiO maser emission exhibit a round appearance with a slight elongation, which agrees with the simulation results shown in \citet{chiong22}. Both the horizontal and vertical polarization components share a similar shape. To determine the beam size, we fitted the combined SiO ($v=1$) maser intensity of both Pol0 (H) and Pol1 (V) components, and took their mean.
The Half Power Beam Width (HPBW) at 43 GHz is measured to be 38.8\arcsec $\times$ 36.6\arcsec. Additionally, a sidelobe is visible at a level of about 1\% of the peak intensity.
This result is in excellent agreement with that shown in \citet{chiong22}, where the beam pattern was obtained in the 2021--2022 season.
We note that the average of the last three year measurements is $\sim$ 38.5\arcsec $\times$ 37\arcsec.

\subsection{Beam efficiencies}

In April 2022, we conducted measurements of the aperture and main beam efficiencies by observing Jupiter. The LO frequency was set to 37 GHz using the NRO 5--7 GHz band continuum backend system. For the observations, we adopt reference frequencies of 31 GHz (LSB - Lower Sideband) and 43 GHz (USB - Upper Sideband).
Taking into account the apparent diameter of Jupiter, which was measured to be 35\arcsec
\footnote{We refer to the value reported by the NAOJ, \url{https://eco.mtk.nao.ac.jp/koyomi/index.html.en}}, along with the obtained peak intensity $T_A^*$, we derived the aperture efficiency ($\eta_A$) and main beam efficiency ($\eta_m$) for both frequencies. 
We also assumed the brightness temperature to be 150 K.
For 31 GHz, the derived values were ($\eta_A$, $\eta_m$)=(0.61, 0.75), while for 43 GHz, the values were (0.61, 0.73).

These efficiency values are consistent with the measurements obtained from other receivers, such as H40 and Z45, at 43 GHz, considering the associated uncertainties (errors) in the measurements. Additional information regarding these measurements can be found in the NRO website\footnote{\url{https://www.nro.nao.ac.jp/~nro45mrt/html/prop/eff/eff_latest.html}}. For example, the main beam efficiency of Z45 is reported to be 0.72 at 43 GHz \citep{nakamura15}.

\begin{center}
 \begin{table*}
  \caption{Important Spectral Transitions in the Q-band}
 \begin{tabular}{lcccc}
 \hline
 Species & Frequency & transition & $E_u$ & Note \\ 
  & (GHz) & & (K) &   \\ \hline
SO & 30.0015 & $J_N$ = $1_0$--$0_1$ & 1.43984 & shock, dense gas, Zeeman \\ 
SO & 36.2018 & $J_N$ = $2_3$--$2_2$ & 21.05 & shock, dense gas, Zeeman \\ 
HC$_5$N & 31.9518 & $J$ = 12--11 &9.96735 &  dense gas\\ 
HC$_5$N & 34.6144 & $J$ = 13--12 & 11.62858 & dense gas \\ 
CCS & 33.7514& $J_N$ = $3_2$--$2_1$ & 3.22575 & early phase, dense gas, Zeeman \\ 
CCS & 45.3790 & $J_N$ = $4_3$--$3_2$ & 5.40356 & early phase, dense gas, Zeeman \\ 
C$^{34}$S & 48.2069 & $J$ = 1--0 & 2.31355 & dense gas \\
$^{13}$CS & 46.2475 & $J$ = 1--0 & 2.2195 & dense gas \\
CS & 48.9900 & $J$ = 1--0 & 2.35118 & dense gas 
\\ 
SiO & 43.4238 & $J$ = 1--0 & 2.08400 & shock \\ 
CH$_3$OH & 44.9558 & $2_0$--$3_1$ E $v_t=1$ &  299.624 & warm gas \\ 
CH$_3$OH & 48.2476 & $1_0$--$0_0$ E $v_t=1$ & 294.993 & warm gas\\ 
CH$_3$OH & 48.2573 & $1_0$--$0_0$ A$^+$ $v_t=1$ & 425.963 & warm gas \\ 
CH$_3$OH & 48.3725 & $1_0$--$0_0$ A$^+$ $v_t=0$ & 2.322 &  \\ 
CH$_3$OH & 48.3769 & $1_0$--$0_0$ E $v_t=0$ & 7.547 & \\
CH$_3$OH & 36.1693 & 4($-$1,4)--3(0,3) E & 28.789 & maser, class I, high-mass protostars \\ 
CH$_3$OH & 44.0693 & 7(0,7)--6(1,6) A$^{++}$ & 64.981 & maser, class I, high-mass protostars \\
H51$\alpha$ -- H59$\alpha$ &  31.2233--48.1537  &     &  &  HII regions
\\ \hline
 \end{tabular}
 \label{tab:line}
\end{table*}
\end{center}

\section{Important Spectral Lines in the Q Band}
\label{sec:line}

The molecular transitions within the Q band offer a unique opportunity to study the star formation process, spanning from its early stages within parent molecular clouds, through the prestellar core and first core phases, to the birth of protostars.

{\color{black} Firstly, a critical but not well constrained factor in the star formation study is the strengths of magnetic fields, which could play important roles in the gas dynamics.
A natural and crucial step for resolving this problem is to accurately measure the strength of magnetic fields associated with prestellar cores, although gauging the strength of interstellar magnetic fields has proven to be challenging.}

To overcome this challenge, the Q-band spectrum presents an invaluable resource. It contains three Zeeman lines of different molecular species, including SO ($J_N=1_0-0_1$), CCS ($J_N=3_2-2_1$), and CCS ($J_N=4_3-3_2$). 
These lines exhibit critical densities of $10^4$ cm$^{-3}$, making them particularly well-suited for tracing dense cores in molecular clouds. Moreover, these lines have large Land\'{e} factors: 
CCS ($J_N=4_3-3_2$) at 45 GHz, CCS ($J_N=3_2-2_1$) at around 34 GHz, and SO ($J_N=1_0-0_1$) at roughly 30 GHz.
The presence of a 100 $\mu$Gauss magnetic field is anticipated to induce a frequency split on the order of $10^2$ Hz in these spectral lines. Importantly, the eQ receiver serves as an excellent tool in this measurement as it has the unique capability for simultaneous Zeeman observations of all these three lines. 

The eQ is also suited for standard spectral line observations of molecular and atomic transitions. 
For example, the lowest rotational transition of the diatomic molecule CS (carbon monosulfide), with a rest frequency of 48.99 GHz, is one of the important dense gas tracers. Emissions from CS and its isotopologues, such as $^{13}$CS and C$^{34}$S, are commonly employed as tracers for high-density regions of $10^3-10^4$ cm$^{-1}$ 
\citep{gardner78,tatematsu93,tsuboi99}.

In addition to CS, carbon-chain molecules like HC$_3$N, HC$_5$N, and CCS are invaluable tracers for similar density regimes. The abundances of these carbon-chain molecules evolve over time during the evolution of molecular clouds and cores \citep{suzuki92,wolkovitch97,lai00,marka12,nakamura14, taniguchi16, shimoikura18,shimoikura19}.
The presence and abundance of carbon-chain molecules are thus useful diagnostics for probing the chemical processes associated with the formation and evolution of dense regions in the interstellar medium. Particularly, these molecules are excellent tracers of prestellar cores, the birthplaces of protostars.

When a prestellar core undergoes gravitational collapse, it results in the formation of a dense adiabatic core at its center, commonly referred to as the first core or the first hydrostatic core \citep{larson69}. The density within the first core exceeds $10^{10}$ cm$^{-3}$, spanning sizes of a few astronomical units and maintaining temperatures in the range of a few hundred Kelvin. In such warm environments, torsionally-excited $v_t=1$ CH$_3$OH lines, with an upper energy level of around 300 K, offer promising prospects as tracers for the first core. The observations of these specific CH$_3$OH lines provide valuable insights into the physical properties and dynamics of the first core during its initial evolutionary stages. Moreover, the presence CH$_3$OH, a precursor of complex organic molecules (COMs), is tightly related to the formation of COMs in space \citep{oberg09}, which is essential to our understanding of the origins of life in the universe.

Once star formation begins, it leads to high-velocity mass ejections known as molecular outflows and jets, accompanied by their interactions with surrounding gas that generate shocks. 
The collision of high-speed gas flows generates intense heating and compression, leading to the release of molecules, including SO, CH$_3$OH, and SiO (silicon monoxide) from interstellar dust particles.
Therefore, their abundances increase significantly through these shock interactions
\citep{haschick90,schlike97,bachiller01,handa06}. Within the Q-band, the $^3\Sigma$ SO $J_N=1_0-0_1$ transition at 30 GHz and the SiO $J=1-0$ transition at a 43 GHz  can be used as shock tracers.
These shock tracers also are proven invaluable for detecting shock-associated events such as dense core collisions \citep{kinoshita22,yano24} and cloud-cloud collisions \citep{nguyen13,nakamura12a,dobashi19b,kinoshita21b}.
CH$_3$OH (class I) maser lines at 36 GHz and 44 GHz emit from outflow-interacting spots around high-mass protostars, and can be used as excellent tracers of high-mass star-forming cores \citep{matsumoto14,rodriguez17,Voronkov14}. 

In addition to the molecular lines introduced above, the Q-band encompasses several hydrogen recombination lines, from H51$\alpha$ (at 48.15 GHz) to H59$\alpha$  (at 31.22 GHz). These lines are exceptional tracers for studying stellar feedback processes and the interaction between clouds and the intense UV radiation emitted by massive stars \citep{nguyen17}.

\begin{center}
 \begin{table*}
  \caption{SBC observations}
 \begin{tabular}{ccccc}
 \hline
 Type & Number of scans & total on-integration time & total observing time & Noise level achieved \\ \hline
Standard & 14 & 4m40s & $\sim$ 15m & 0.1524 K \\ 
SBC1 & 17 & 5m40s & $\sim$ 15m & 0.0951 K \\ 
SBC2 & 14 & 4m40s &$\sim$ 10m &  0.1033 K\\ 
SBC3 & 7 & 2m20s & $\sim$ 6m & 0.1446 K \\ \hline
 \end{tabular}
 \label{tab:sbc}
\end{table*}
\end{center}

\section{Smoothed Bandpass Calibration}
\label{sec:sbc}

\begin{figure}[htbp]
 \begin{center}
   \includegraphics[width=\linewidth]{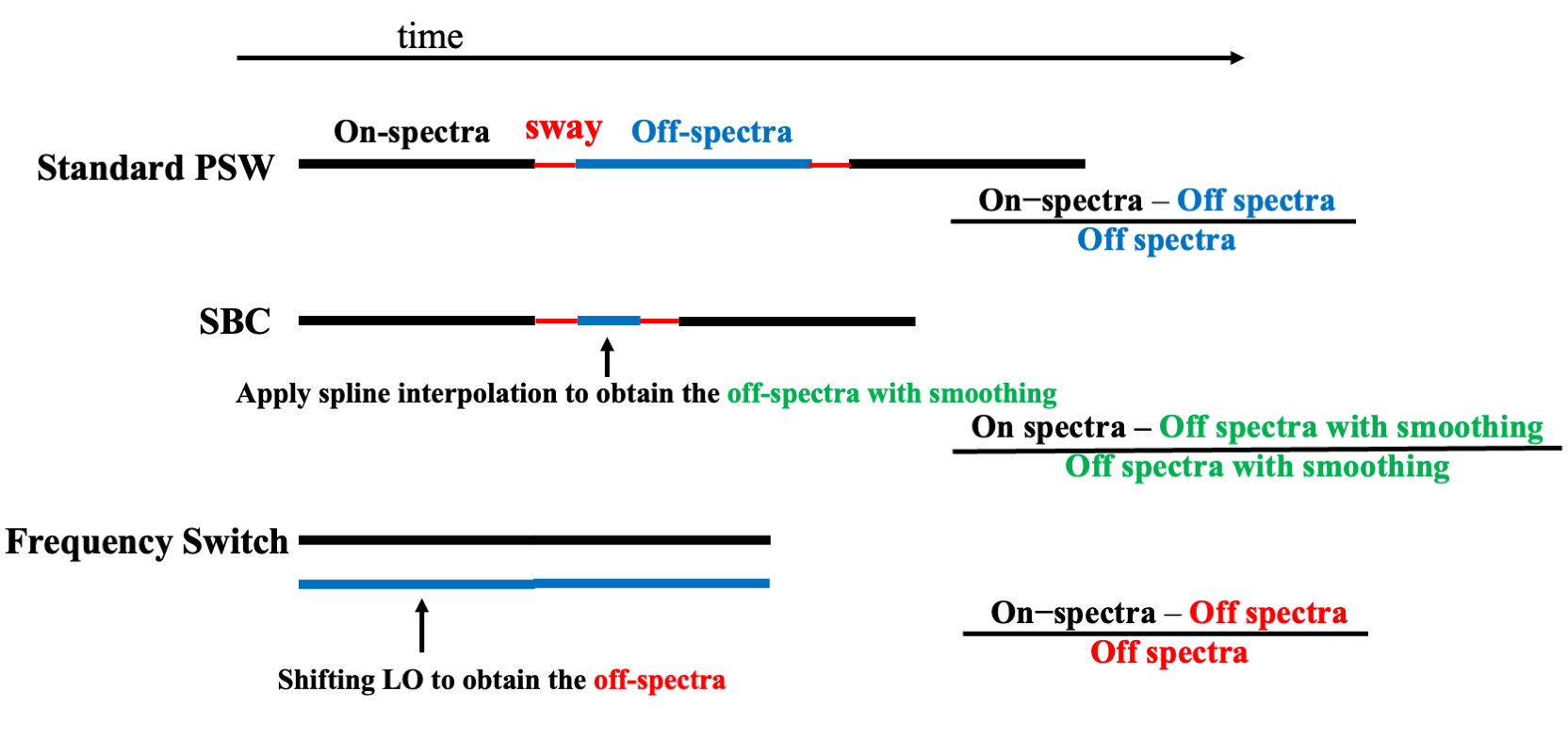}
 \end{center}
\caption{Comparison of three observation methods, (1) Standard Position-SWitch, (2) Smoothed Bandpass Calibration, and (3) Frequency Switch.
The black, red, and blue lines indicate the ON-position integration time, telescope sway time (between ON and OFF positions), and the OFF position integration time, respectively.
For PSW and SBC, the telescope is physically moved between ON and OFF positions (sway time). For Frequency switch, the emission-free spectra is obtained by changing the LO frequency without moving the telescope.
}
\label{fig:obscomparison}
\end{figure}

\begin{figure*}[htbp]
 \begin{center}
   \includegraphics[width=\linewidth]{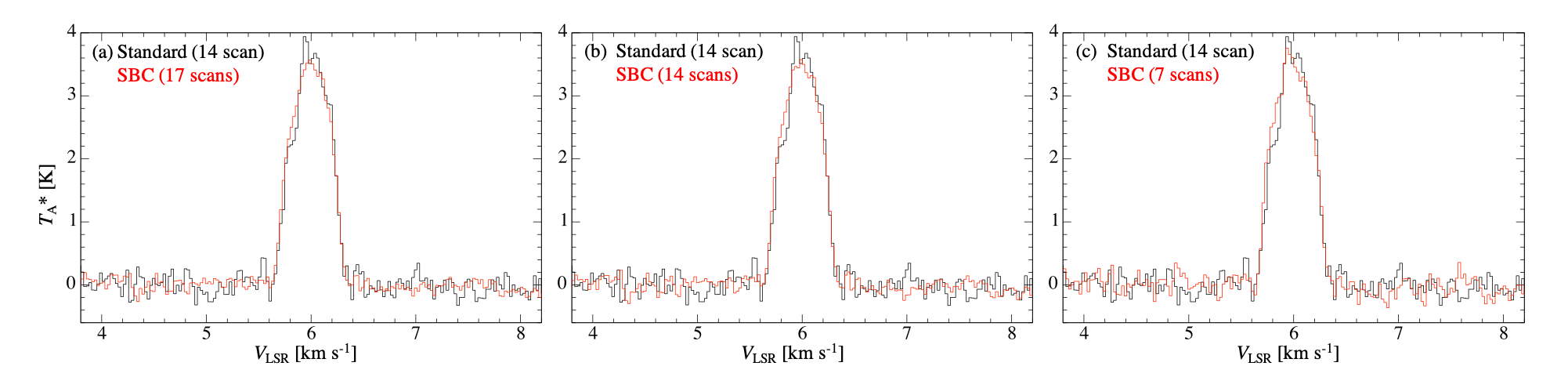}\hspace{1cm}
 \end{center}
\caption{Comparison of the line profiles between the standard observations and SBC technique. Panels (a) through (c) indicate the comparison with the line profile obtained by the standard on-off observations, (a) the same total observation time case, (b) the same integration time case, and (c) the same noise level case. The target line is HC$_5$N ($J=17-16$). The black lines are the profiles obtained by the standard observations, while the red lines are the profiles obtained with SBC.
}
\label{fig:sbc}
\end{figure*}

As described above, the eQ receiver has higher-sensitivity, making it particularly suitable for detecting faint emission lines and acquiring data with excellent signal-to-noise ratios in a limited observation time.
To further enhance observational efficiency, we employ the smoothed bandpass calibration (SBC) technique, as introduced by \citet{yamaki12}.

In the standard on-off observations using a position switch, the integration time for the emission-free (off) position is matched with that of the target position. In contrast, the SBC method optimizes the integration time for the emission-free position by applying data smoothing techniques to the spectra.

Another switching technique, known as frequency switch, acquires emission-free spectra by shifting the LO frequency while maintaining the same integration time. 
This technique also shortens the total observation time compared to the standard position switch observations, although this is not suitable for broad and/or crowded spectral lines.

Figure \ref{fig:obscomparison} presents a comparative analysis of these three methods.
The actual reduction efficiency in the SBC method is assessed through the stability of the receiver Allan variance \citep{yamaki12}.
For example, in the case of the previous receiver, Z45, our experiment indicated that 
a configuration with 120 seconds on-source integration and 10 seconds off-source integration 
offers optimal efficiency with 256 channel smoothing.
This configuration was subsequently adopted for the Zeeman observations \citep{nakamura19}, resulting in a three times reduction in total observation time.
Previous investigations have successfully applied the SBC method, yielding substantial enhancements in signal-to-noise ratios \citep{taniguchi16,taniguchi18,dobashi18,nakamura19}.  Taking into account these past experiences, we anticipate a reduction in total observation time 
by a factor of 2.5 -- 3 through the implementation of the SBC method.
Therefore, the observational efficiency of the SBC method is comparable to that of the frequency switch and more efficient than the standard position-switch observations.

Here, we demonstrate effectiveness of the SBC method in significantly minimizing total observation time.
Based on our previous experiments, we configured the on-source and off-source observation durations to 20 seconds and 5 seconds, respectively, for the SBC observations, and
implemented 32 channel-smoothing for the off-source spectra using the cubic spline interpolation.
These SBC parameters were determined as follows; we adjusted the off-integration time, shortening it while keeping the on-integration time constant.
Then, we tested two smoothing cases of 16 and 32 channels which correspond to about 0.4 and 0.8 km s$^{-1}$ velocity resolutions, respectively. The adopted smoothing channel-width is unlikely to significantly affect the baseline determination.
The narrow bandwidth of our observations, i.e., those focused on Galactic objects, minimizes artificial undulation caused by instrumental and atmospheric origins. 
In addition, we confirmed the noise level is somewhat smaller with 32 channel-smoothing, as expected. 
These specific SBC parameters have been well tested and successfully employed in previous Nobeyama observations \citep{taniguchi16,taniguchi18}.
More optimized parameters should be determined based on the Allan variance measurements. This is presumably more significant for high-redshift object observations with wider bandwidth.

The observations were conducted in December 2023. The target position is TMC-1 (CP). The wind speed was below 1 m s$^{-1}$. The system noise temperatures at about 45 GHz were approximately 140 K and did not change during the observations. 
We created the following three spectra applied with the SBC; 
(SBC1) maintaining the same total observation time including both on- and off-source observations, (SBC2) preserving the same on-source integration time (4 min 40 sec, or 14 scans), 
and (SBC3) achieving noise levels similar to that observed in non-SBC spectra. 
The specifics of these observations are briefly outlined in Table \ref{tab:sbc}.

Figures \ref{fig:sbc}(a) through \ref{fig:sbc}(c) present comparisons between the line profiles obtained using the standard ON-OFF observations ({\it black}) and the SBC method ({\it red}), respectively. 

These profiles indicate that the profiles obtained by the SBC coincide well with that of the standard on-off observations.
To achieve the same noise level as the standard observations (SBC3), the SBC reduces the total observation time by a factor of 2.5. 
This result agrees with those of the previous observations \citep{taniguchi16,taniguchi18}.

\section{Rest Frequencies of CCS ($J_N=4_3-3_2$) and SO ($J_N=1_0-0_1$)}
\label{sec:obs1}

One of our primary sciences is to conduct Zeeman observations with CCS ($J_N=4_3-3_2$), CCS ($J_N=3_2-2_1$), and SO ($J_N=1_0-0_1$).
In Table \ref{tab:freq}, we have compiled the rest frequencies of these molecular lines, 
which highlight the presence of small but non-negligible discrepancies between different catalogs, where we adopted the frequency values from Splatalogue\footnote{\url{https://splatalogue.online//advanced1.php}}. 
At lower frequencies, a deviation of around 10 kHz translates to a velocity difference of $\sim$ 0.1 km s$^{-1}$, roughly half the sound speed at $T=10$ K.
This becomes particularly relevant in quiescent environments, where typical starless cores exhibit narrow line widths approaching the sound speed.
For example, when considering the CCS ($J_N=3_2-2_1$), CCS ($J_N=4_3-3_2$), 
and SO ($J_N=1_0-0_1$) transitions, we observe frequency differences of 4.1 kHz, 17 kHz, 
and 56.5 kHz, respectively (taking the maximum minus the minimum values). 
They correspond to the velocity differences of 0.03, km s$^{-1}$, 0.11 km s$^{-1}$, 
and 0.44 km s$^{-1}$, respectively. Particularly, the SO ($J_N=1_0-0_1$) line has a huge difference, comparable to the typical linewidth, and it is therefore crucial to accurately constrain the rest frequencies of these molecular lines.
Such deviations impede a comprehensive understanding of cloud structure. For example, the TMC-1 contains many sub-filaments with velocity differences typically around 0.1 km s$^{-1}$ \citep{dobashi19}. Identifying corresponding components becomes challenging 
when using different lines due to these discrepancies (see Figure \ref{fig:so_tmc1cp}).

\begin{center}
 \begin{table}
  \caption{CCS and SO rest frequencies listed in Splatalogue $^*$. 
}
 \begin{tabular}{ccc}
 \hline
 Rest frequency (kHz) & Error(kHz) & Catalog \\ [0.5ex] \hline
 CCS $J_N=3_2-2_1$ &  &  \\ [0.5ex] 
 33751369.6 & 1.0 &  CDMS \\ 
 33751370.0 & 40.0 & Lovas \\ 
 33751370.0 & 4.0 & SLAIM \\ 
 33751373.7 & 0.8 & JPL \\ 
 \hline 
 CCS $J_N=4_3-3_2$ &  &  \\ [0.5ex] 
 45379046.0 & 20.0 & CDMS \\ 
 45379029.0 & 2.0 &  Lovas \\ 
 45379029.0 & 5.0 & SLAIM \\ 
 45379033.0 & 1.0 & JPL \\ 
 \hline 
 SO $J_N=1_0-0_1$ &  &  \\ [0.5ex] 
 30001580.0& 20.0 & CDMS \\ 
 30001547.0 & 2.0 &  Lovas \\ 
 30001547.0 & 5.0 & SLAIM \\ 
 30001523.5 & 1.0 & JPL \\ 
   \hline
 \end{tabular}
 \label{tab:freq} \\
 $^*$ Lovas (NIST Recommended Rest Frequencies), 
  SLAIM (Spectral Line Atlas of Interstellar Molecules), 
  JPL (Jet Propagation Laboratory Molecular Spectroscopy),
  CDMS (The Cologne Database for Molecular Spectroscopy)
\end{table}
\end{center}

To improve our ability in analyzing the observational data for better interpreting the underlying physical process,
it is imperative to achieve a more precise determination of the rest frequencies of these Zeeman transitions. 
Given the high sensitivity of the eQ receiver, it becomes relatively easy to obtain spectral lines with a high signal-to-noise ratio within a short observation time. Thus, it allows us to directly compare the frequencies of these transitions from the observations. 


\subsection{Tragets, methods, and CCS rest frequencies}
\label{subsec:freq}

We have selected three distinct target positions for our comparative analysis, as listed in Table \ref{tab:target}.
Our primary reference position is TMC-1 (CP). Previous observations have demonstrated that the CCS emission lines at this location exhibit pronounced intensity and distinctively sharp line edges (i.e., the derivatives of the spectral line, $dI/dv$, are large near the line edges). These characteristics make TMC-1 (CP) an excellent choice for conducting this measurement, even though there are multiple molecular components overlapping along the line-of-sight.

In addition to TMC-1 (CP), we have designated two supplementary target positions, the ammonia peak of TMC-1 and Polaris Flare, to corroborate the results obtained from TMC-1 (CP). 
The Polaris Flare cloud is a diffuse cloud containing a starless dense clump (MCLD123.5+24.9), and the CCS and HC$_3$N emission with narrow line widths are detected in an earlier study  \citep{shimoikura12}. This property makes Polaris Flare an ideal candidate for cross-referencing and validating our results.

\begin{center}
 \begin{table*}
  \caption{Targets}
 \begin{tabular}{lccccc}
 \hline
 Name & R.A. & Dec. & $V_{\rm LSR}$ & off-position & Note \\ 
   &  &  & km s$^{-1}$ & ($\alpha, \delta$) &  \\\hline
TMC-1(CP)& 04h41m43.87s & +25:41:17.7 & 5.85 & (30\arcmin, 0\arcmin) & \citet{dobashi18} \\ 
TMC-1(NH$_3$)& 04h41m34.58s & +25:45:50.1 & 5.85 & (30\arcmin, 0\arcmin) &  \\ 
Polaris Flare & 01h58m58.51s & +87:39:46.7 & $-$4.3 & (30\arcmin, 0\arcmin) &  \citet{shimoikura12} \\ 
   \hline
 \end{tabular}
 \label{tab:target}
\end{table*}
\end{center}

The procedure for constraining the rest frequencies is briefly outlined as follows.
\begin{itemize}
\item[1.] We observe the three Zeeman lines simultaneously with the frequency resolution of 3.81 kHz, the finest frequency resolution of the NRO SAM45 spectrometer, toward all three target regions.
\item[2.] We adopt the CCS ($J_N$ = $3_2$--$2_1$) line as the reference, since the differences in the rest frequencies among the catalogs are small (within $\sim $ 1 channel of SAM45).
\item[3.] We compare the CCS ($J_N$ = $4_3$--$3_2$) line with the reference line, and attempt to match the CCS ($J_N$ = $4_3$--$3_2$) line's velocity axis to that of $J_N$ = $3_2$--$2_1$ toward the TMC-1 (CP) sightline.
\item[4.] We further compare the SO line with the CCS ($J_N$ = $3_2$--$2_1$) transition observed again toward the same TMC-1 (CP) sightline.
\item[5.] Finally, using the rest frequencies determined from the above procedure, we reconfirm whether the line profiles are consistent with each other at the other two sightlines.
\end{itemize}

Observations were done in September 2023. 
The telescope pointing was determined by observing a SiO maser source, NML Tau (KL Tau), and the typical pointing offset was better than 3\arcsec \ during the whole observing period.
The standard ON-OFF observations with the ON-source integration time of 20 sec were carried out using the NRO digital spectrometer SAM45, providing the highest spectral resolution of 3.81 kHz.  We conducted simultaneous observations of six molecular lines: HC$_3$N ($J$ = 5--4), DC$_3$N ($J$ = 5--4), CCS ($J_N$ = $4_3$--$3_2$), HC$_5$N ($J$ = 19--18), CCS ($J_N$ = $3_2$--$2_1$), and SO ($J_N$ = $1_0$--$0_1$). However, in this paper, we specifically present the profiles of three selected lines: CCS ($J_N$ = $4_3$--$3_2$), CCS ($J_N$ = $3_2$--$2_1$), and SO ($J_N$ = $1_0$--$0_1$).   We note that the other lines also have similar rest frequency discrepancy among the catalogs. 
The emission-free position at ($\Delta$ R.A., $\Delta$ Dec.)=(30\arcmin, 0\arcmin) was adopted for the OFF-point observations. The total on-source time was 40 min.
The achieved rms noise levels are 81 mK and 53 mK in $T_A^*$ for the $J_N$ = $4_3$--$3_2$ and $J_N$ = $3_2$--$2_1$ transitions, respectively.

We chose the CCS ($J_N$ = $3_2$--$2_1$) transition as the reference line since the frequency difference (4.1 kHz) among the four catalogs within about 1 channel of the SAM45 spectrometer, and therefore we assumed that the rest frequency measurement is most reliable.  Since the average frequency is calculated to be 33.7513708 GHz, we adopt the same value as that of Lovas and SLAIM catalogs (33.751370 GHz) as the CCS ($J_N$ = $3_2$--$2_1$) rest frequency.  

\begin{figure}[htbp]
 \begin{center}
    \includegraphics[angle=0, width=\linewidth]{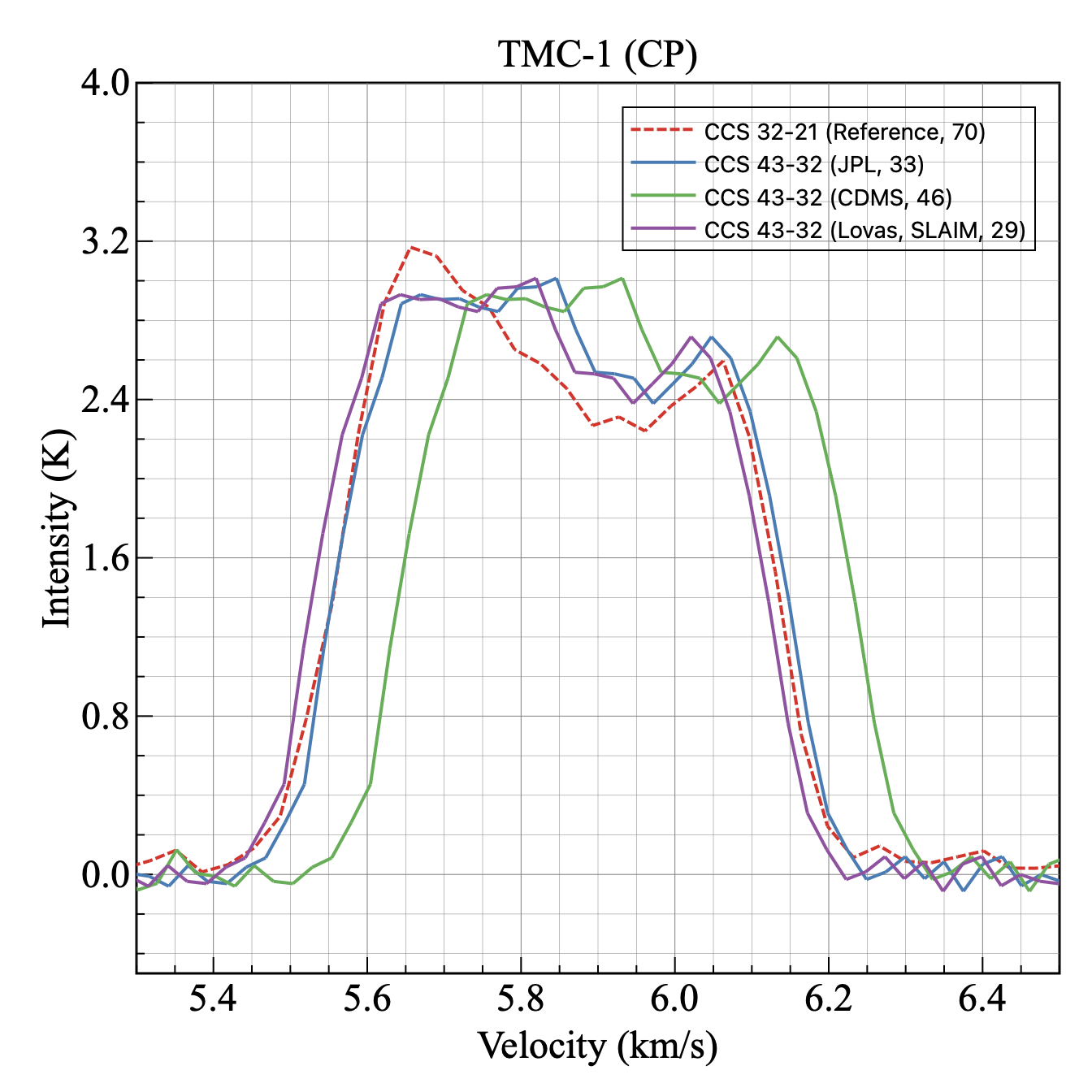}\hspace{1cm}
 \end{center}
\caption{Comparison between CCS ($J_N$ = $3_2$--$2_1$) and CCS ($J_N$ = $4_3$--$3_2$) line profile at TMC-1 (CP).
The intensities are given in T$_A^*$ (K).
In the Legend, we show the information of the catalog name and the last two digits of the rest frequencies in kHz.
The red dashed line is the reference spectral line, CCS ($J_N$ = $3_2$--$2_1$), with the rest frequency of 33751370.0 kHz (Lovas, SLAIM). The blue, green, and purple solid lines are the CCS ($J_N$ = $4_3$--$3_2$) with JPL, CDMS, Lovas/SLAIM rest frequencies, respectively. The JPL value shows the best agreement with the reference profile.  A smaller rest frequency makes the profile shift toward the left or smaller velocity.}
\label{fig:tmc-1freq}
\end{figure}

In Figure \ref{fig:tmc-1freq}, we compare the two CCS spectral lines observed at TMC-1 (CP). 
This figure indicates that the CCS ($J_N$ = $4_3$--$3_2$) line with the rest frequency in the JPL catalog is in best agreement with the reference line, the CCS ($J_N$ = $3_2$--$2_1$) transition.

\begin{figure*}[htbp]
 \begin{center}
    \includegraphics[angle=0, width=0.45 \linewidth]{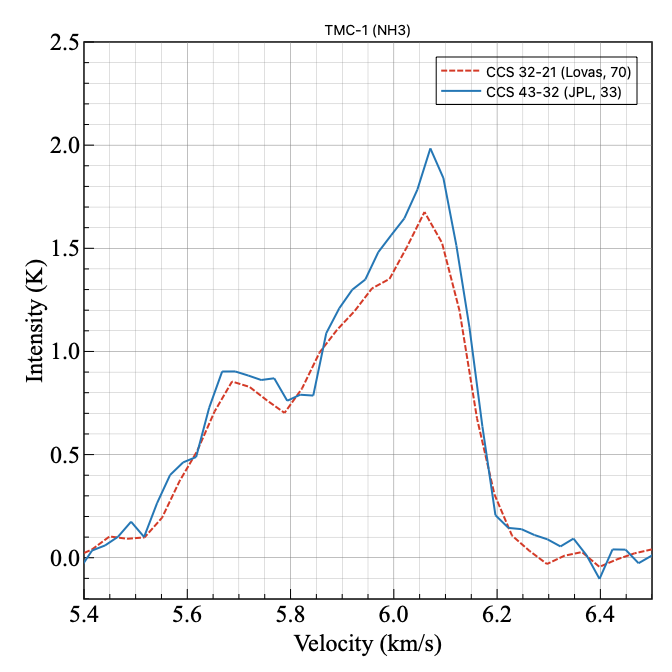}
     \includegraphics[angle=0, width=0.45 \linewidth]{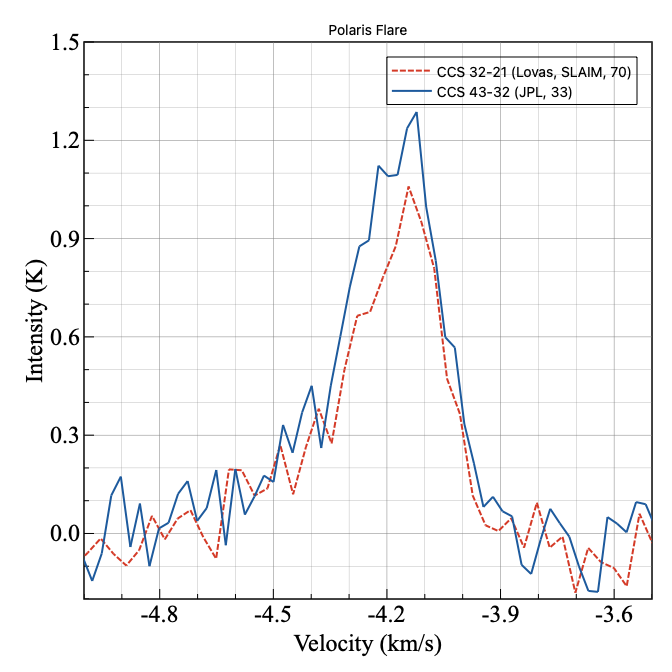}
 \end{center}
\caption{The CCS ($J_N$ = $3_2$--$2_1$) and CCS ($J_N$ = $4_3$--$3_2$) line profiles toward (a) TMC-1 (NH$_3$) and (b) Polaris Flare.
The red dashed line is the reference spectral line, CCS ($J_N$ = $3_2$--$2_1$),
while the solid line is the CCS ($J_N$ = $4_3$--$3_2$) line with the JPL rest frequency.
The rms noise levels are measured to be ($J_N$ = $4_3$--$3_2$, $J_N$ = $3_2$--$2_1$) = (83 mK, 39 mK) and (159 mK, 125 mK) for TMC-1 (NH$_3$) and Polaris, respectively.
}
\label{fig:ccs_comparison}
\end{figure*}

Next, we compare the CCS ($J_N$ = $4_3$--$3_2$) profile based on the JPL frequency with the CCS ($J_N$ = $3_2$--$2_1$) transition with the reference frequency observed at the other two positions.  
Figures \ref{fig:ccs_comparison}(a) and \ref{fig:ccs_comparison}(b) show such comparisons toward TMC-1 (NH$_3$) and Polaris Flare, respectively.
The spectral profile against velocity for both targets exhibit an excellent match with each other.
Therefore, we adopt 45.379033 GHz as the rest frequency of CCS ($J_N$ = $4_3$--$3_2$).

There are two important considerations to acknowledge in this analysis. First, the beam sizes vary for lines at different frequencies. This discrepancy in spatial resolution can potentially impact the observed line shapes and intensity distributions. Second, it is likely that  some of the observed lines are optically thick, with optical depths estimated to be around $\tau \sim$ 1--2. For example, according to \citet{dobashi18}, the components A to D of the CCS ($J_N$ = $4_3$--$3_2$) line at the TMC-1 (CP) have the optical depths of $\tau \sim$ 2.2--1.1.
This optical thickness introduces additional complexities and may influence the line profiles.
These caveats should be taken into account when interpreting the results, as the observed line shapes and intensities may be influenced by the combination of beam size variations and optical thickness effects. However, in the present paper, we did not take into account these effects for simplicity.

\subsection{Rest frequency of SO ($J_N$ = $1_0$--$0_1$)}

\begin{figure}[htbp]
 \begin{center}
    \includegraphics[angle=0, width= \linewidth]{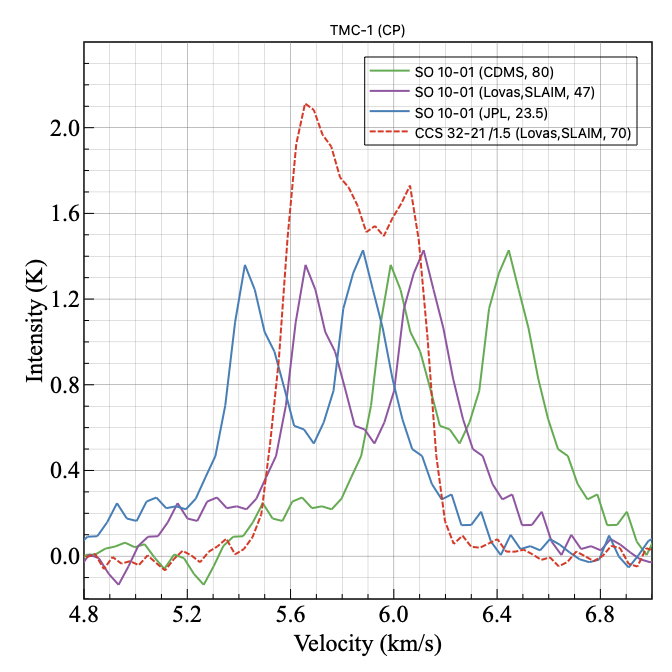}
 \end{center}
\caption{Comparison between the SO ($J_N$ = $1_0$--$0_1$) and CCS ($J_N$ = $3_2$--$2_1$) line profiles toward TMC-1 (CP).
The rest frequencies indicated in the legend is used to draw the SO line profiles. 
The red dashed line is the reference spectral line, CCS ($J_N$ = $3_2$--$2_1$).
The rms noise level of SO is measured to be 83 mK.
}
\label{fig:so_tmc1cp}
\end{figure}

The rest frequency of SO ($J_N$ = $1_0$--$0_1$) is not well constrained, having a discrepancy of $\sim$ 50 kHz and leading to $\sim$ 0.4 km s$^{-1}$ velocity difference.  
Here, we assume that the SO line is emitted from similar regions where the CCS lines are emitted.

We first compare the reference line CCS ($J_N$ = $3_2$--$2_1$) and the SO lines with the three different rest frequencies. 
The analysis of SO ($J_N=1_0-0_1$) presents a more complex situation due to its double-peaked structure, and therefore the comparison is not as simple.
Figure \ref{fig:so_tmc1cp} shows the reference line profile and the SO lines with different three rest frequencies, indicating that the overlap with the reference line looks the best for the Lovas/SLAIM SO profile.
Upon visual inspection, the two peaks appear to correspond to the two overlapping sub-components near the line edges, referred to as components A and D identified with CCS ($J_N$ = $4_3$--$3_2$) in \citet{dobashi18}, which showed that the CCS profile consists of four components with different velocities of (A) 5.64, (B) 5.82, (C) 6.00 and (D) 6.08 km s$^{-1}$.
We note that \citet{dobashi18} adopted the CDMS rest frequency, and therefore we modified the velocities with JPL's.
\citet{dobashi18} determined these velocities using the excellent frequency resolution data of 60 Hz which corresponds to 40 cm s$^{-1}$, and therefore the 
velocities of the sub-components are very accurate.
To match the two peaks of SO, the rest frequency should be equal to 30.001542 GHz, slightly smaller (5 kHz) than the Lovas/SLAIM value.
For this frequency, the SO line takes its peaks at 5.61 km s$^{-1}$ and 6.07 km s$^{-1}$.

It is worth noting that the CH$_3$OH ($1_0$--$0_0$ A$^+$ at 48.3724558 GHz) profile shows a similar two-peak shape at 5.6 km s$^{-1}$ and 5.9 km s$^{-1}$ \citep{soma18}.
These are in good agreement with the velocities of the SO ($J_N$ = $1_0$--$0_1$) line.
Therefore, we conclude that the best rest frequency of SO is likely to be 
very close to the Lovas/SLAIM value.
Our best value we infer from the comparison between the CCS and SO profiles is 30.001542 GHz. It is worth noting that CH$_3$OH lines also have uncertainty of the rest frequencies.

\begin{figure*}[htbp]
 \begin{center}
    \includegraphics[angle=0, width= 0.45\linewidth]{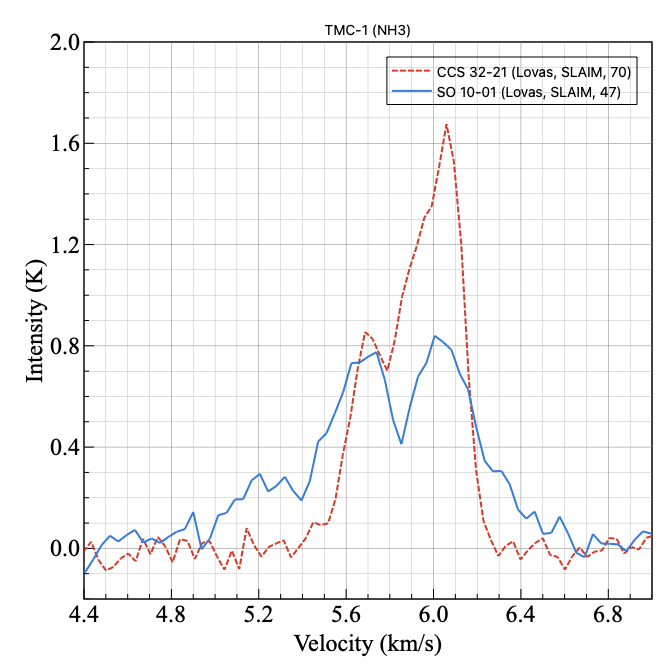}
        \includegraphics[angle=0, width= 0.45\linewidth]{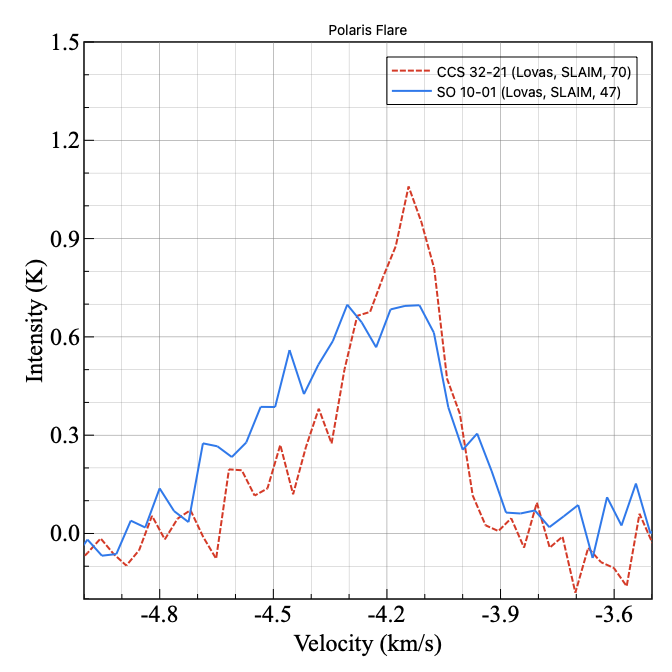}
 \end{center}
\caption{Comparison between SO ($J_N$ = $1_0$--$0_1$) and CCS ($J_N$ = $3_2$--$2_1$) line profiles toward ({\it left}) TMC-1 (CP) and ({\it right}) Polaris Flare.
The intensities are given in T$_A^*$ (K).
The rest frequencies indicated in the legend is used to draw the SO line profiles. 
The red dashed line is the reference spectral line, CCS ($J_N$ = $3_2$--$2_1$).
The rms noise level of SO is measured to be 83 mK.
}
\label{fig:so_comparison}
\end{figure*}

In Figure \ref{fig:so_comparison}, we compare the SO profile with the Lovas/SLAIM value with the reference line profile for the two positions.  Although the detailed line shape is different from that of the reference line, the extents of the SO profile are in excellent agreement with that of the reference CCS line for both positions.

\section{Taurus Molecular Cloud-1}
\label{sec:obs2}

\subsection{Mapping observations toward TMC-1}

\begin{figure*}[htbp]
 \begin{center}
   \includegraphics[angle=0, width=8cm]{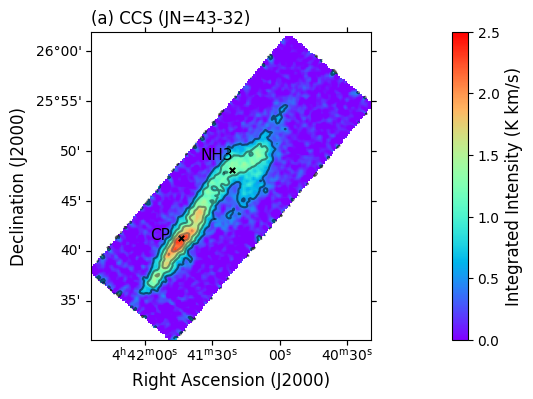}
      \includegraphics[angle=0, width=8cm]{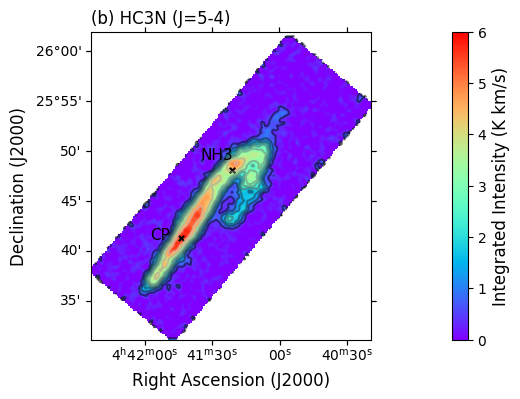}
      \includegraphics[angle=0, width=8cm]{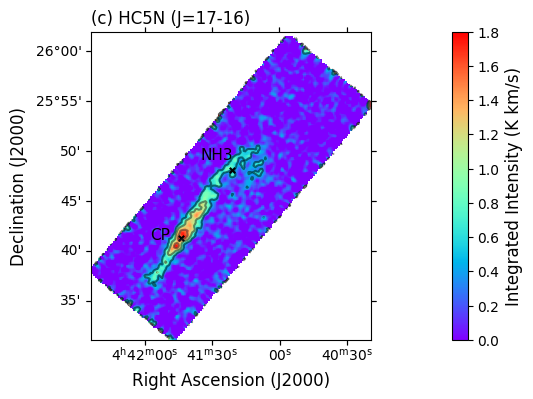}
            \includegraphics[angle=0, width=8cm]{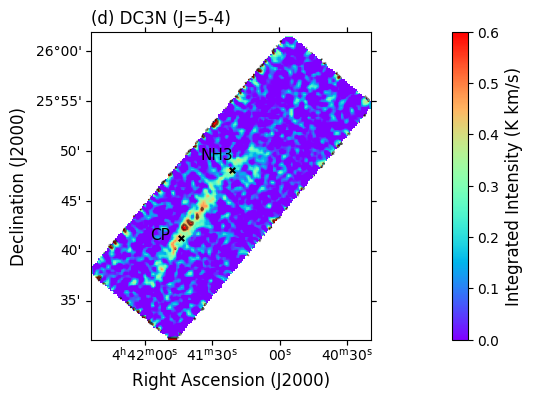}
\includegraphics[angle=0, width=8cm]{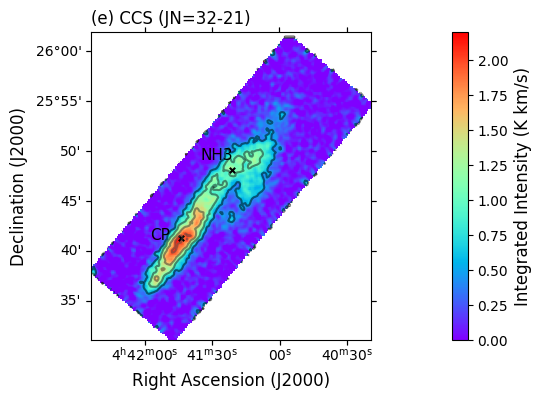}
\includegraphics[angle=0, width=8cm]{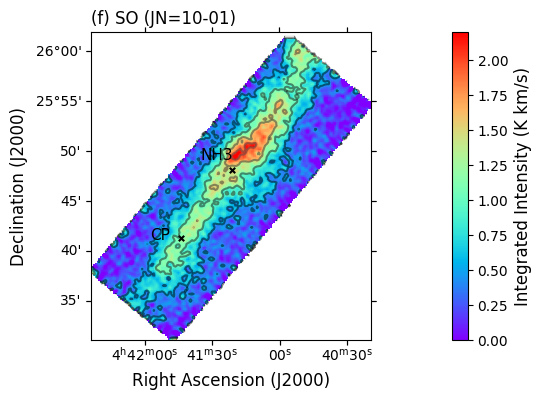}
 \end{center}
\caption{Velocity-integrated intensity maps of 
(a) CCS ($J_N$ = $4_3$--$3_2$), (b) HC$_3$N ($J$ = 5--4), (c) HC$_5$N ($J$ = 17--16), (d) DC$_3$N ($J$ = 5--4), (e) CCS ($J_N$ = $3_2$--$2_1$),
and (f) SO ($J_N$ = $1_0$--$0_1$) toward TMC-1.
The intensities are integrated in the range of 4.5 km s$^{-1}$ to 7.0 km s$^{-1}$, given in K km s$^{-1}$.
The intensities are corrected by the main beam efficiencies of 0.75 and 0.73 for 43--45 GHz [CCS ($J_N$ = $4_3$--$3_2$), HC$_3$N ($J$ = 5--4), HC$_5$N ($J$ = 17--16), DC$_3$N ($J$ = 5--4)], and 30--34 GHz [CCS ($J_N$ = $3_2$--$2_1$), SO ($J_N$ = $1_0$--$0_1$)], respectively.
The positions of TMC1 (CP) and TMC1 (NH$_3$) are indicated by crosses.
}
\label{fig:tmc1map}
\end{figure*}

In Figure \ref{fig:tmc1map}, we present integrated-intensity maps of several molecular emission lines toward TMC-1, a prestellar filament. 
Panels (a) though (f) of Figure \ref{fig:tmc1map} show the 
(a) CCS ($J_N$ = $4_3$--$3_2$), (b) HC$_3$N ($J$ = 5--4), (c) HC$_5$N ($J$ = 17--16), (d) DC$_3$N ($J$ = 5--4), (e) CCS ($J_N$ = $3_2$--$2_1$),
and (f) SO ($J_N$ = $1_0$--$0_1$) velocity integrated maps, respectively.
The six molecular lines were obtained simultaneously with a 3.81 kHz frequency resolution ($\simeq$ 0.025 km s$^{-1}$ at 45 GHz). The observations were done in a OTF mode and the typical $T_{\rm sys}$ were about 100 K and 60 K for 45 GHz and 33 GHz, respectively.
The pointing observations were done with SiO maser lines from NML-Tau (KL Tau) every 1--1.5 hours, and pointing errors were within 3--5\arcsec.

The spatial distributions of carbon-chain molecules, CCS and HC$_3$N, exhibit remarkable similarities, sharing common distribution characteristics.
The filament traced by these carbon-chain molecules is running from southeast to northwest. 
They are more intense in the southern part of the TMC-1 filament.
In contrast, the SO emission is stronger in the northern part of the TMC-1 filament.
The distribution appears anti-correlated with those of the carbon-chain molecules.
This trend is consistent with the SO ($J_N$ = $3_2$--$2_1$), 93 GHz observations made by \citet{hirahara95}.
The different distribution patterns reflect distinct chemical evolutionary phases experienced by the northern and southern regions of TMC-1.
Furthermore, our analysis has unveiled the presence of multiple components coexisting along the line of sight for both CCS and SO.
For example, at the position of ammonia peak, the two local peaks at around 5.7 km s$^{-1}$ and 6.0 km s$^{-1}$ appear to match with each other (compare Figures \ref{fig:ccs_comparison} and \ref{fig:so_comparison}) 
These components appear to agree with velocity-coherent subfilaments within the overarching filamentary structure, as previously identified by \citet{dobashi19} using CCS ($J_N$ = $4_3$--$3_2$). We will discuss these features in the futher paper.

\subsection{Implication for the evolution at TMC-1 (CP)}

The SO and CH$_3$OH are released from dusts by shock heating process. Therefore,  similar shapes of these molecular lines may come from the shock heating event.  Based on the extremely high spectral resolution (60 Hz, corresponding to 40 cm s$^{-1}$ velocity resolution) observations,  \citet{dobashi18} revealed that the CCS profile consists of the four components at the CP position. By solving the line-of-sight radiation-transfer calculations, they determined the most plausible spatial configuration of the four components (which is A (5.64 km s$^{-1}$)-B (5.82 km s$^{-1}$)-C (6.00 km s$^{-1}$)-D (6.08 km s$^{-1}$) in the order of distance along the TMC-1 (CP) sightline), and that A and B components are approaching toward us and the other two are moving away. In other words, 4 components are infalling toward the center ($\approx$ 5.96 km s$^{-1}$) at this position. 
\citet{nakamura19} conducted the Zeeman observations and measured $\sim 117 \mu$G strength, and suggested that this position is magnetically supercritical with the mass-to-flux ratio of 2.2, even if the plane-of-sky component estimated from the near infrared linear polarization map is taken into account. The infall speeds of A and D are also about three times the sound speed, which is consistent with the 
 terminal velocity of the gravitational infalling isothermal core \citep{larson69}. 
Therefore, the CP position is likely to be gravitationally-infalling toward the center \citep{nakamura19}. 
The strong SO emission at A and D components
is consistent with this interpretation when
the gravitationally-driven motion generates the shock at the outer part of the filament (A and D).

As mentioned above, the similar multi-peak profiles can be found at various positions along the main filament \citep{dobashi19}. For example, at the ammonia peak, the CCS lines show two peaks at $\sim$ 5.7 km s$^{-1}$ and $\sim$ 6.0 km s $^{-1}$.  The SO also peaks at similar velocities.
These profiles may be due to the gravitational collapse driven shocks.
In a separate paper, we will discuss the cloud structure of TMC-1 in more details.

\section{High-z objects}

For the high-$z$ universe, the CO ($J$ = 1--0), CO ($J$ = 2--1), and CO ($J$ = 3--2) lines emitted from $z$ = 1.30--2.83, 3.60--6.67, 
and 5.92--10.52, respectively, fall onto our bandwidth, since our universe is expanding. The observed frequency at a redshift $z$ becomes smaller as
\begin{equation}
\nu_{\rm obs} = {\nu_{\rm rest} \over 1 + z}
\end{equation}
where $\nu_{\rm obs}$ is the observed redshifted frequency, and $\nu_{\rm rest}$ is the rest frequency of the line. 
CO ($J$ = 1--0), CO ($J$ = 2--1), and CO ($J$ = 3--2) have $\nu_{\rm rest}$ = 115.271202 GHz, 230.53800 GHz, and 345.7959899 GHz, respectively.
These low-$J$ CO lines serve as excellent tracers of the total molecular gas content. Previous observations conducted with the GBT have successfully detected CO ($J$ = 1--0) lines using several hours of integration time. Considering the increased sensitivity of our new receiver, we anticipate being able to detect these CO lines in a comparable integration time.

Here, we demonstrate the capabilities of our eQ receiver by observing a previously CO-detected high-z object.
We chose a strongly-lensed Hyper Luminous Infrared Galaxy, PJ120207.6 
at a redshift of 2.442.
The parameters of this target are listed in Table \ref{tab:high-z}.
For the observations, we set the center frequency to 33.497 GHz, following \citet{harrington18}. During the observations, the typical system noise temperature was about 60--80 K.


\begin{center}
 \begin{table*}
  \caption{High-$z$ Target}
 \begin{tabular}{lccccc}
 \hline
 Name & R.A. & Dec. & center frequency (GHz) & $z$ & Note \\ 
  PJ120207.6 (PLCK G138.6+62.0) & 12h02m07.6s & +53d34m39s & 33.4970 & 2.442 & \citet{harrington18} \\
   \hline
 \end{tabular}
 \label{tab:high-z}
\end{table*}
\end{center}

\begin{figure}[htbp]
 \begin{center}
    \includegraphics[angle=0, width=\linewidth]{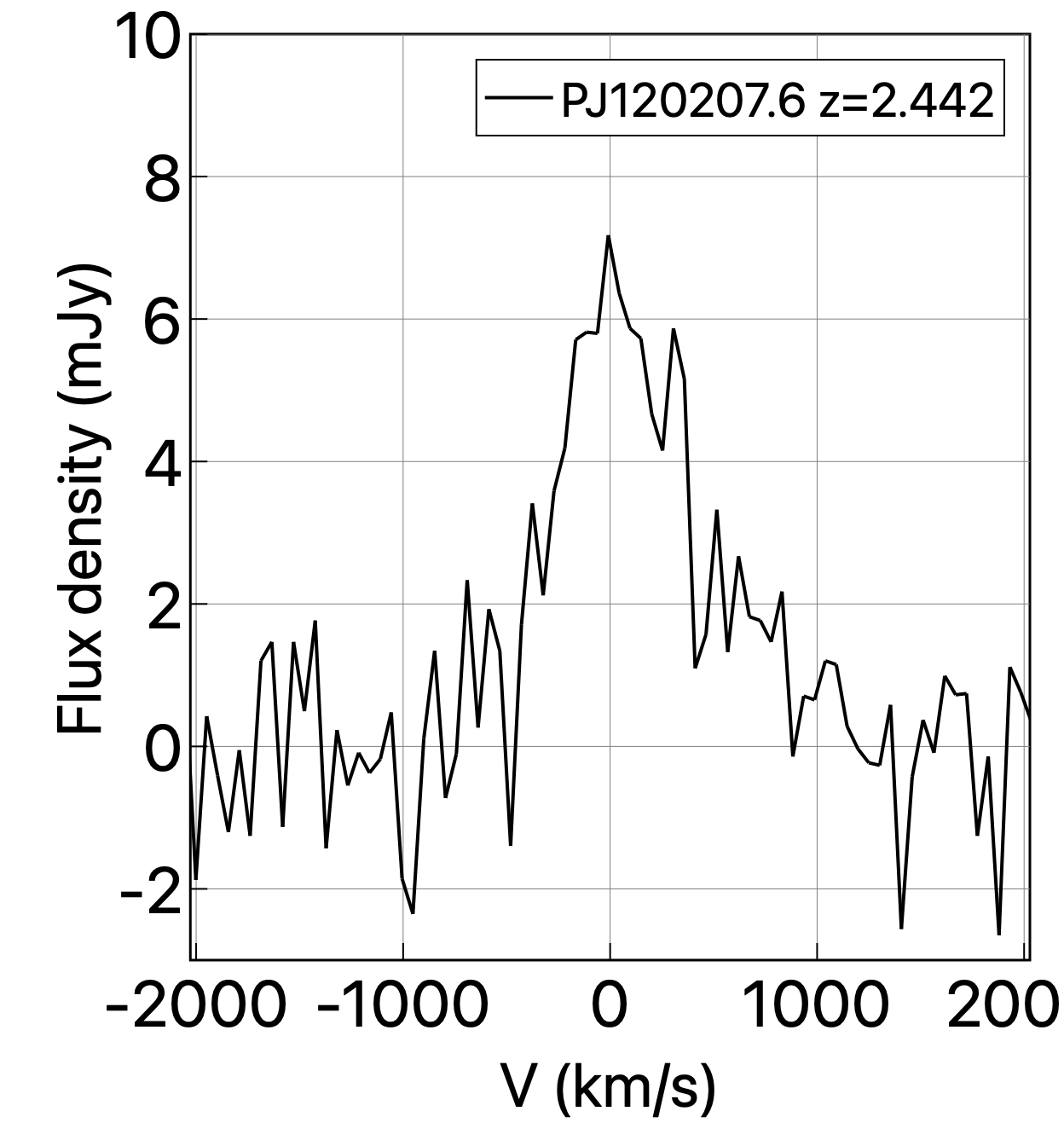}
 \end{center}
\caption{The CO ($J=1-0$) spectrum  of PJ120207.6.}
\label{fig:high-z}
\end{figure}

We successfully detected the subtle CO emission originating from PJ 120207.6, which is shown in Figure \ref{fig:high-z}. 
The total on-source integration time was 2 hours. Including the observations of the emission-free position, the total observation time was about 4--4.5 hours.
The data were acquired using a 2 GHz mode, with a frequency resolution of 488.28 kHz. To set the velocity resolution of 50 km s$^{-1}$, we aggregated and smoothed the spectral profile.
The peak intensity was about 4 mK and we converted it to Jy by using the conversion factor of 0.754 K/Jy adopting the beam size of 38\arcsec and the beam efficiency of 0.75.
The peak value is consistent with that of \citet{harrington18}.
The line shape is more or less similar to the CO ($J$ = 1--0) and CO ($J=3-2$)  profiles \citep{canameras15, harrington18}.
    

\section{Summary}
\label{sec:summary}
We have developed a cutting-edge Q-band receiver for single dish telescopes, named  eQ, in collaboration with ASIAA, Taiwan, and NAOJ, Japan. In this paper, we provide a detailed account of the receiver's specifications, its installation on the Nobeyama 45-m telescope in November 2021, and present the results of commissioning observations. The key findings and insights from this work can be summarized as follows:
\begin{itemize}
\item[1.] 
The eQ receiver is a state-of-the-art dual linear polarization Q-band receiver having a wide bandwidth of 30 to 50 GHz and excellent sensitivity. It exhibits a receiver noise temperature of about 15 K over the 30 to 50 GHz range. During installation in the Nobeyama 45-m telescope, we measured system noise temperatures of 30 K and 70 K for frequencies of 33 and 45 GHz, respectively, under excellent sky conditions.
\item[2.]
Our primary scientific goals with the eQ receiver encompass: (1) conducting Zeeman observations using CCS and SO to measure the magnetic field strengths of dense molecular cloud cores, (2) detecting molecular lines emitted from high-redshift objects to determine their total molecular gas mass and redshifts, and (3) exploring astrochemistry in the Q-band.

\item[3.]
We demonstrated the effectiveness of the SBC technique for the eQ observations.  
We successfully reduced the total observation time by a factor of about 2.5.

\item[4.]  
We conducted simultaneous observations of target Zeeman lines of CCS ($J_N$ = $4_3$--$3_2$), CCS ($J_N$ = $3_2$--$2_1$), and SO ($J_N$ = $1_0$--$0_1$), employing a frequency resolution of 3.81 kHz. We compared the obtained spectral lines and attempted to refine their rest frequencies. Particularly, the rest frequency of SO had not been precisely determined, showing discrepancies of about 50 kHz from existing catalogs. Our observations led us to determine the rest frequencies of CCS ($J_N$ = $4_3$--$3_2$) and SO ($J_N$ = $1_0$--$0_1$) as 45.79033 GHz and 30.001542 GHz, respectively, using the CCS ($J_N$ = $3_2$--$2_1$) value of 33.751370 GHz as a reference.

\item[5.]  We conducted observations targeting TMC-1, a nearby dense filament. The SO spectral line exhibited double peaks at the CP position and other positions within TMC-1. Comparing these profiles with the CCS profile consisting of four components revealed a strong correspondence between the SO peaks and the CCS components situated in the outer part of the filament. 
According to \citet{dobashi18} and \citet{nakamura19}, the CP position is magnetically supercritical and undergoes gravitational contraction. 
If this is the case, the SO double peaks 
are likely to originate from shock compression induced by gravitational contraction, given that SO tends to be enhanced by shock heating. 
The emission of SO is more pronounced in the northern part of the filament, while CCS is more prominent in the southern part. In simpler terms, the distribution of SO and CCS along the filament appears to be opposite to each other, suggesting distinct chemical processes taking place in different regions of the filament.
\item[5.] We successfully detected the CO ($J$ = 1--0) emission line from the high-redshift galaxy PJ12027.6 at a redshift of $z$ = 2.442.
\end{itemize}


\begin{ack}
This work was financially supported by JSPS KAKENHI Grant Numbers JP23H01218 (F.N.), JP22K02966 (T. S.), JP20K14523 (K.T.), JP21H01142 (K. T.), and 2022 research fund of NAOJ Division of Science.
N.H. acknowledge support from the National Science and Technology Council of Taiwan (NSTC) with grant NSTC 111-2112-M-001-060.
T.S. has been financially supported by the Kayamori Foundation of Informational Science Advancement.
We thank Shigehisa Takakuwa and Ross A. Burns for their encouragements and supports.
We thank the NRO staff for both operating the 45 m and helping us with the data reduction.
\end{ack}

\bibliographystyle{pasj}
\bibliography{nakamura.bib}

\begin{thebibliography}{45}
\providecommand{\natexlab}[1]{#1}

\bibitem[{{Bachiller} et~al.(2001){Bachiller}, {P{\'e}rez Guti{\'e}rrez},
  {Kumar}, \& {Tafalla}}]{bachiller01}
{Bachiller}, R., {P{\'e}rez Guti{\'e}rrez}, M., {Kumar}, M.S.N., \& {Tafalla},
  M. 2001, \aap, 372, 899

\bibitem[{{Ca{\~n}ameras} et~al.(2015)}]{canameras15}
{Ca{\~n}ameras}, R., et~al. 2015, \aap, 581, A105

\bibitem[{{Chiong} et~al.(2018){Chiong}, {Chen}, {Chang}, {Huang}, \&
  {Hwang}}]{chiong18}
{Chiong}, C.C., {Chen}, C., {Chang}, C.C., {Huang}, Y.D., \& {Hwang}, Y.J.
  2018, in Asia-Pacific Microwave Conference (APMC), ed. IEEE, 18419991

\bibitem[{{Chiong} et~al.(2021){Chiong}, {Chen}, {Ho}, {Jian}, \&
  {Hwang}}]{chiong21}
{Chiong}, C.C., {Chen}, C., {Ho}, C.T., {Jian}, H.Y., \& {Hwang}, Y.J. 2021, in
  50th European Microwave Conference (EuMC), ed. IEEE, 20405111

\bibitem[{{Chiong} et~al.(2022)}]{chiong22}
{Chiong}, C.C., et~al. 2022, in Millimeter, Submillimeter, and Far-Infrared
  Detectors and Instrumentation for Astronomy XI, eds. J.~{Zmuidzinas} \& J.R.
  {Gao}, vol. 12190 of Society of Photo-Optical Instrumentation Engineers
  (SPIE) Conference Series, 121900M

\bibitem[{{Dobashi} et~al.(2019{\natexlab{a}}){Dobashi}, {Shimoikura},
  {Katakura}, {Nakamura}, \& {Shimajiri}}]{dobashi19b}
{Dobashi}, K., {Shimoikura}, T., {Katakura}, S., {Nakamura}, F., \&
  {Shimajiri}, Y. 2019{\natexlab{a}}, \pasj, 71, S12

\bibitem[{{Dobashi} et~al.(2018){Dobashi}, {Shimoikura}, {Nakamura}, {Kameno},
  {Mizuno}, \& {Taniguchi}}]{dobashi18}
{Dobashi}, K., {Shimoikura}, T., {Nakamura}, F., {Kameno}, S., {Mizuno}, I., \&
  {Taniguchi}, K. 2018, \apj, 864, 1, 82

\bibitem[{{Dobashi} et~al.(2019{\natexlab{b}}){Dobashi}, {Shimoikura},
  {Ochiai}, {Nakamura}, {Kameno}, {Mizuno}, \& {Taniguchi}}]{dobashi19}
{Dobashi}, K., {Shimoikura}, T., {Ochiai}, T., {Nakamura}, F., {Kameno}, S.,
  {Mizuno}, I., \& {Taniguchi}, K. 2019{\natexlab{b}}, \apj, 879, 2, 88

\bibitem[{{Gardner} \& {Whiteoak}(1978)}]{gardner78}
{Gardner}, F.F., \& {Whiteoak}, J.B. 1978, \mnras, 183, 711

\bibitem[{{Handa} et~al.(2006){Handa}, {Sakano}, {Naito}, {Hiramatsu}, \&
  {Tsuboi}}]{handa06}
{Handa}, T., {Sakano}, M., {Naito}, S., {Hiramatsu}, M., \& {Tsuboi}, M. 2006,
  \apj, 636, 1, 261

\bibitem[{{Harrington} et~al.(2018)}]{harrington18}
{Harrington}, K.C., et~al. 2018, \mnras, 474, 3, 3866

\bibitem[{{Haschick} \& {Ho}(1990)}]{haschick90}
{Haschick}, A.D., \& {Ho}, P.T.P. 1990, \apj, 352, 630

\bibitem[{{Hirahara} et~al.(1995){Hirahara}, {Masuda}, {Kawaguchi}, {Ohishi},
  {Ishikawa}, {Yamamoto}, {Takano}, \& {Kaifu}}]{hirahara95}
{Hirahara}, Y., {Masuda}, A., {Kawaguchi}, K., {Ohishi}, M., {Ishikawa}, S.I.,
  {Yamamoto}, S., {Takano}, S., \& {Kaifu}, N. 1995, \pasj, 47, 845

\bibitem[{{Hwang} et~al.(2017)}]{hwang17}
{Hwang}, Y.J., et~al. 2017, in 12th European Microwave Integrated Circuits
  Conference (EuMIC), ed. IEEE, 17452024

\bibitem[{{Kinoshita} \& {Nakamura}(2022)}]{kinoshita22}
{Kinoshita}, S.W., \& {Nakamura}, F. 2022, \apj, 937, 2, 69

\bibitem[{{Kinoshita} et~al.(2021)}]{kinoshita21b}
{Kinoshita}, S.W., et~al. 2021, \pasj, 73, S300

\bibitem[{{Lai} \& {Crutcher}(2000)}]{lai00}
{Lai}, S.P., \& {Crutcher}, R.M. 2000, \apjs, 128, 1, 271

\bibitem[{{Larson}(1969)}]{larson69}
{Larson}, R.B. 1969, \mnras, 145, 271

\bibitem[{{Marka} et~al.(2012){Marka}, {Schreyer}, {Launhardt}, {Semenov}, \&
  {Henning}}]{marka12}
{Marka}, C., {Schreyer}, K., {Launhardt}, R., {Semenov}, D.A., \& {Henning}, T.
  2012, \aap, 537, A4

\bibitem[{{Matsumoto} et~al.(2014)}]{matsumoto14}
{Matsumoto}, N., et~al. 2014, \apjl, 789, 1, L1

\bibitem[{{Mizuno} et~al.(2014)}]{mizuno14}
{Mizuno}, I., et~al. 2014, Journal of Astronomical Instrumentation, 3, 1450010

\bibitem[{{Nakamura} et~al.(2012)}]{nakamura12a}
{Nakamura}, F., et~al. 2012, \apj, 746, 25

\bibitem[{{Nakamura} et~al.(2014)}]{nakamura14}
{Nakamura}, F., et~al. 2014, \apjl, 791, L23

\bibitem[{{Nakamura} et~al.(2015)}]{nakamura15}
{Nakamura}, F., et~al. 2015, \pasj, 67, 117

\bibitem[{{Nakamura} et~al.(2019)}]{nakamura19}
{Nakamura}, F., et~al. 2019, \pasj, 71, S3

\bibitem[{{Nguyen-Lu'o'ng} et~al.(2013)}]{nguyen13}
{Nguyen-Lu'o'ng}, Q., et~al. 2013, \apj, 775, 2, 88

\bibitem[{{Nguyen-Lu'o'ng} et~al.(2017)}]{nguyen17}
{Nguyen-Lu'o'ng}, Q., et~al. 2017, \apjl, 844, 2, L25

\bibitem[{{{\"O}berg} et~al.(2009){{\"O}berg}, {Garrod}, {van Dishoeck}, \&
  {Linnartz}}]{oberg09}
{{\"O}berg}, K.I., {Garrod}, R.T., {van Dishoeck}, E.F., \& {Linnartz}, H.
  2009, \aap, 504, 3, 891

\bibitem[{{Rodr{\'\i}guez-Garza} et~al.(2017){Rodr{\'\i}guez-Garza}, {Kurtz},
  {G{\'o}mez-Ruiz}, {Hofner}, {Araya}, \& {Kalenskii}}]{rodriguez17}
{Rodr{\'\i}guez-Garza}, C.B., {Kurtz}, S.E., {G{\'o}mez-Ruiz}, A.I., {Hofner},
  P., {Araya}, E.D., \& {Kalenskii}, S.V. 2017, \apjs, 233, 1, 4

\bibitem[{{Schilke} et~al.(1997){Schilke}, {Walmsley}, {Pineau des Forets}, \&
  {Flower}}]{schlike97}
{Schilke}, P., {Walmsley}, C.M., {Pineau des Forets}, G., \& {Flower}, D.R.
  1997, \aap, 321, 293

\bibitem[{{Shimoikura} et~al.(2018){Shimoikura}, {Dobashi}, {Nakamura},
  {Matsumoto}, \& {Hirota}}]{shimoikura18}
{Shimoikura}, T., {Dobashi}, K., {Nakamura}, F., {Matsumoto}, T., \& {Hirota},
  T. 2018, \apj, 855, 1, 45

\bibitem[{{Shimoikura} et~al.(2019){Shimoikura}, {Dobashi}, {Nakamura},
  {Shimajiri}, \& {Sugitani}}]{shimoikura19}
{Shimoikura}, T., {Dobashi}, K., {Nakamura}, F., {Shimajiri}, Y., \&
  {Sugitani}, K. 2019, \pasj, 71, S4

\bibitem[{{Shimoikura} et~al.(2012){Shimoikura}, {Dobashi}, {Sakurai},
  {Takano}, {Nishiura}, \& {Hirota}}]{shimoikura12}
{Shimoikura}, T., {Dobashi}, K., {Sakurai}, T., {Takano}, S., {Nishiura}, S.,
  \& {Hirota}, T. 2012, \apj, 745, 2, 195

\bibitem[{{Soma} et~al.(2018){Soma}, {Sakai}, {Watanabe}, \&
  {Yamamoto}}]{soma18}
{Soma}, T., {Sakai}, N., {Watanabe}, Y., \& {Yamamoto}, S. 2018, \apj, 854, 2,
  116

\bibitem[{{Suzuki} et~al.(1992){Suzuki}, {Yamamoto}, {Ohishi}, {Kaifu},
  {Ishikawa}, {Hirahara}, \& {Takano}}]{suzuki92}
{Suzuki}, H., {Yamamoto}, S., {Ohishi}, M., {Kaifu}, N., {Ishikawa}, S.I.,
  {Hirahara}, Y., \& {Takano}, S. 1992, \apj, 392, 551

\bibitem[{{Taniguchi} et~al.(2016){Taniguchi}, {Ozeki}, {Saito}, {Sakai},
  {Nakamura}, {Kameno}, {Takano}, \& {Yamamoto}}]{taniguchi16}
{Taniguchi}, K., {Ozeki}, H., {Saito}, M., {Sakai}, N., {Nakamura}, F.,
  {Kameno}, S., {Takano}, S., \& {Yamamoto}, S. 2016, \apj, 817, 2, 147

\bibitem[{{Taniguchi} et~al.(2018)}]{taniguchi18}
{Taniguchi}, K., et~al. 2018, \apj, 866, 2, 150

\bibitem[{{Tatematsu} et~al.(1993)}]{tatematsu93}
{Tatematsu}, K., et~al. 1993, \apj, 404, 643

\bibitem[{{Tercero} et~al.(2021)}]{tercero21}
{Tercero}, F., et~al. 2021, \aap, 645, A37

\bibitem[{{Tsuboi} et~al.(1999){Tsuboi}, {Handa}, \& {Ukita}}]{tsuboi99}
{Tsuboi}, M., {Handa}, T., \& {Ukita}, N. 1999, \apjs, 120, 1, 1

\bibitem[{{Voronkov} et~al.(2014){Voronkov}, {Caswell}, {Ellingsen}, {Green},
  \& {Breen}}]{Voronkov14}
{Voronkov}, M.A., {Caswell}, J.L., {Ellingsen}, S.P., {Green}, J.A., \&
  {Breen}, S.L. 2014, \mnras, 439, 3, 2584

\bibitem[{{Wolkovitch} et~al.(1997){Wolkovitch}, {Langer}, {Goldsmith}, \&
  {Heyer}}]{wolkovitch97}
{Wolkovitch}, D., {Langer}, W.D., {Goldsmith}, P.F., \& {Heyer}, M. 1997, \apj,
  477, 241

\bibitem[{{Yamaki} et~al.(2012){Yamaki}, {Kameno}, {Beppu}, {Mizuno}, \&
  {Imai}}]{yamaki12}
{Yamaki}, H., {Kameno}, S., {Beppu}, H., {Mizuno}, I., \& {Imai}, H. 2012,
  \pasj, 64, 118

\bibitem[{{Yano} et~al.(2024){Yano}, {Nakamura}, \& {Kinoshita}}]{yano24}
{Yano}, Y., {Nakamura}, F., \& {Kinoshita}, S.W. 2024, arXiv e-prints,
  arXiv:2402.11147

\bibitem[{{Zhong} et~al.(2018)}]{zhong18}
{Zhong}, W.Y., et~al. 2018, Research in Astronomy and Astrophysics, 18, 4, 044

\end{thebibliography}

\end{document}